\documentclass[12pt]{elsarticle}

\usepackage[utf8x]{inputenc}
\usepackage[super]{nth}

\usepackage[hyphens]{url}
\usepackage{hyperref}

\usepackage{multirow}
\usepackage{booktabs} 
\usepackage{graphicx}
\usepackage{subcaption}
\usepackage{comment}
\usepackage{xspace}
\usepackage{color}
\usepackage{flushend}
\usepackage{arydshln}
\usepackage{amsmath}
\usepackage{xcolor}
\usepackage{amssymb}
\usepackage[absolute,overlay]{textpos}

\journal{}

\usepackage[pscoord]{eso-pic}

\begin{document}

\title{Debate on Online Social Networks at the Time of COVID-19: An Italian Case Study}

%% Document starts
\author[1]{Martino Trevisan\corref{cor1}}
\author[1]{Luca Vassio\corref{cor1}}
\author[1]{Danilo Giordano\corref{cor1}}

\address[1]{Politecnico di Torino, Torino, Italy}

\cortext[cor1]{Corresponding authors\\
\textit{Email address:} martino.trevisan@polito.it (Martino Trevisan), luca.vassio@polito.it (Luca Vassio), danilo.giordano@polito.it (Danilo Giordano).}

\TPshowboxestrue
\TPMargin{0.3cm}
\begin{textblock*}{15.5cm}(3cm,1.4cm)
\footnotesize
\bf
\definecolor{myRed}{rgb}{0.55,0,0}
\color{myRed}
\noindent
Please cite this article as: Martino Trevisan, Luca Vassio, Danilo Giordano. Debate on Online Social Networks at the Time of COVID-19: An Italian Case Study. Online Social Networks and Media (2021). DOI: \url{https://doi.org/10.1016/j.osnem.2021.100136}
\end{textblock*}

\makeatletter
\def\ps@pprintTitle{%
 \let\@oddhead\@empty
 \let\@evenhead\@empty
 \def\@oddfoot{}%
 \let\@evenfoot\@oddfoot}

\newcommand\footnoteref[1]{\protected@xdef\@thefnmark{\ref{#1}}\@footnotemark}
\makeatother

\begin{abstract}

The COVID-19 pandemic is not only having a heavy impact on healthcare but also changing people's habits and the society we live in. Countries such as Italy have enforced a total lockdown lasting several months, with most of the population forced to remain at home. During this time, online social networks, more than ever, have represented an alternative solution for social life, allowing users to interact and debate with each other. Hence, it is of paramount importance to understand the changing use of social networks brought about by the pandemic.
In this paper, we analyze how the interaction patterns around popular influencers in Italy changed during the first six months of 2020, within Instagram and Facebook social networks. We collected a large dataset for this group of public figures, including more than $54$ million comments on over $140$ thousand posts for these months.
We analyze and compare engagement on the posts of these influencers and provide quantitative figures for aggregated user activity. We further show the changes in the patterns of usage before and during the lockdown, which demonstrated a growth of activity and sizable daily and weekly variations. We also analyze the user sentiment through the psycholinguistic properties of comments, and the results testified the rapid boom and disappearance of topics related to the pandemic. To support further analyses, we release the anonymized dataset.

\end{abstract}

\begin{keyword}
Social Network; Big Data; COVID-19; Facebook; Instagram.
\end{keyword}

\maketitle

\color{black}
\section{Introduction}
\label{sec:intro}

\begin{figure}[t]
    \centering
    \includegraphics[width=0.7\columnwidth]{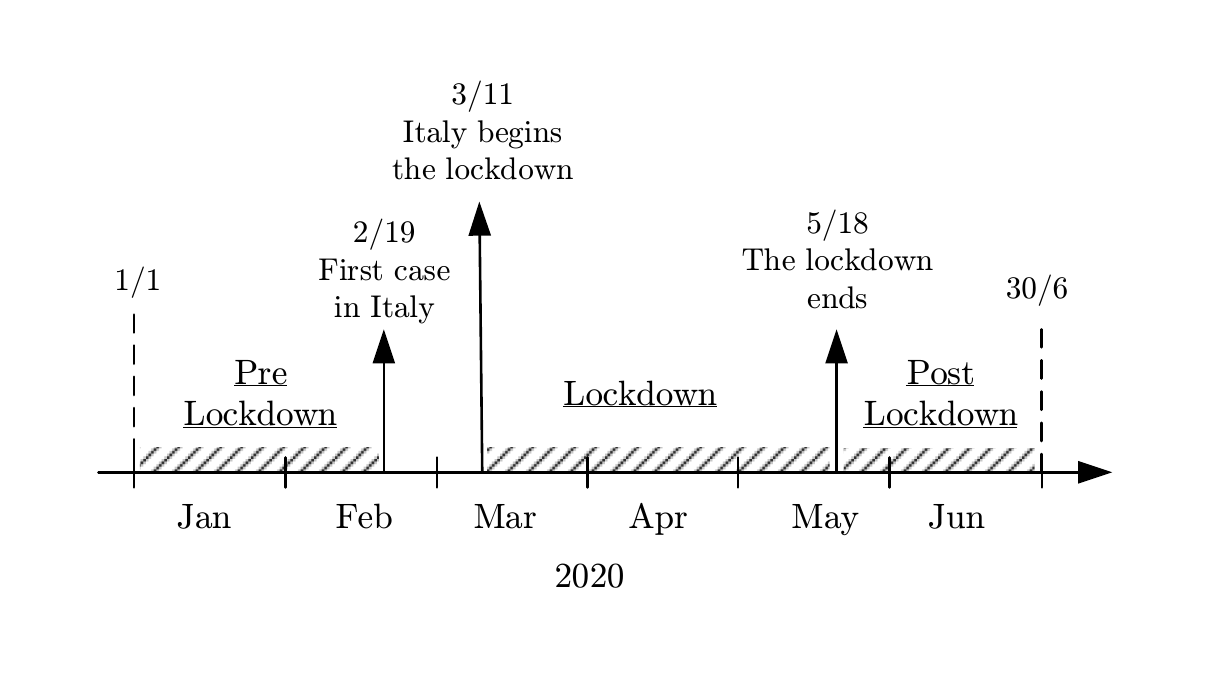}
    \caption{Main events related to the COVID-19 outbreak in Italy in the first six months of 2020.}
    \label{fig:events}
\end{figure}

The COVID-19 pandemic is having a massive impact on people's lives and habits around the world. The countries most affected by the virus are facing an unprecedented health crisis, whose effects are impacting society, economy, culture and politics. Italy was among the first countries hit by COVID-19. The first case was identified on February 19$^{th}$, and two days later the Government issued the first law decree to impose quarantine in the limited area where the disease had broken out. On February 25$^{th}$, due to the alarming growth of cases the Government imposed remote working for all public offices and shut down schools and classes at Universities, in four regions in the north of Italy. Finally, on March 11$^{th}$, the ``\#IoRestoACasa'' decree imposed a total lockdown throughout Italy. People were only allowed to leave the house for valid and proven reasons. Common retail businesses, catering and restaurant services were suspended, gatherings in public places were prohibited. As a consequence, Italy entered the most restrictive lockdown in its history. Restrictions were only slightly relaxed on May 4$^{th}$, when people were allowed to leave the house to practice sport individually or to meet relatives and close family friends. On May 18$^{th}$, the lockdown officially ended, people were allowed to leave the house and many public activities reopened. To guide the reader through our analyses, Figure~\ref{fig:events} shows the timeline of the main events related to the outbreak in Italy during the first six months of 2020.

In general, by interacting with each other through posts, comments, and alike, users build complex networks that favor the dissemination of information~\cite{Al-Garadi:2018}. Understanding how users interact with each other on these platforms is thus of paramount importance to understand how the online debate impacts our society\cite{Conover:2012, Gorkovenko:2017, Pierri:2020}, especially during this historical event.  The lockdown restrictions limited people's mobility and therefore Internet applications like online collaboration platforms, e-learning, gaming, and video streaming rapidly increased their popularity. Correspondingly, Internet traffic volume grew by about $40\%$~\cite{cloudflare}. In this context, social media applications represented an alternative way to physical meetings and social life. 

In this paper, we analyze the changes in users' interaction patterns around popular influencers in Italy on two popular social networks: Instagram and Facebook. Facebook (FB) is currently established as the world's most popular social media application, while the popularity of Instagram (IG) has surged in recent years~\cite{tstat5years,Whashington:2020}. On Facebook and Instagram, \emph{profiles} follow \emph{influencers'} (popular public profiles) and can {\it like}/{\it react}/{\it comment} on their posts. We base our study on a dataset consisting of $54.8$ $million$ comments over $140$ $thousands$ posts written by $639$ influencers in Italy. The dataset covers a period of six months before and after the first lockdown in Italy, from January $1^{st}$ 2020 to July $1^{st}$ 2020. We monitored selected influencers to include heterogeneous categories of profiles i.e., athletes, entertainers, musicians, and politicians. We tracked all posts of each influencer over the period, recording all comments, likes, reactions, and commenters associated with those posts. 

We provide quantitative figures on the impact of the COVID-19 pandemic on social network behaviour in Italy. We analyzed and compared user engagement and participation before, during, and after the lockdown, studying the trends of activity patterns, interactions, and engagement in discussions about specific topics. We group our investigations under the following research questions, limited to popular Italian influencers:

\begin{itemize}
    \item How did the use of Facebook and Instagram and interaction patterns change among users in Italy before, during, and after the lockdown? (Section~\ref{sec:charact}, Section~\ref{sec:4.1}, Section~\ref{sec:4.3}, and Section~\ref{sec:4.4})
    \item  How did the psycholinguistic properties of comments and the topics which users discussed vary in the two social networks? (Section~\ref{sec:4.5} and Section~\ref{sec:4.6})
\end{itemize}

Our results show an increase in social network usage during the lockdown. Facebook led this increase; we also observed a shift in user habits in general, with Italian users more active during the morning and on Friday and Saturday evenings in the period studied. In the same period, the number of followers of Instagram political profiles increased dramatically. Analysis of psycholinguistic properties and topics demonstrate how, during the lockdown, comments expressing concerns such as anxiety and inhibitions increased, as did the level of discussion around topics related to the pandemic. We believe this paper provides useful results regarding changes in discussions on social networks around popular influencers during the COVID-19 outbreak in Italy. These results may help other researchers and sociologists who study human behavior. To this end, we have made our anonymized dataset available online~\cite{ourwebsite}.

The remainder of this article is organized as follows: Section~\ref{sec:dataset-methods} presents our methods for gathering and processing the data. In Section~\ref{sec:results}, we present our results. In Section~\ref{sec:discusion} our results are discussed and placed within the context of related work. Finally, Section~\ref{sec:conclusions} concludes the paper. 
\section{Dataset and methods}
\label{sec:dataset-methods}

\begin{figure}[t]
    \centering
    \includegraphics[width=\columnwidth]{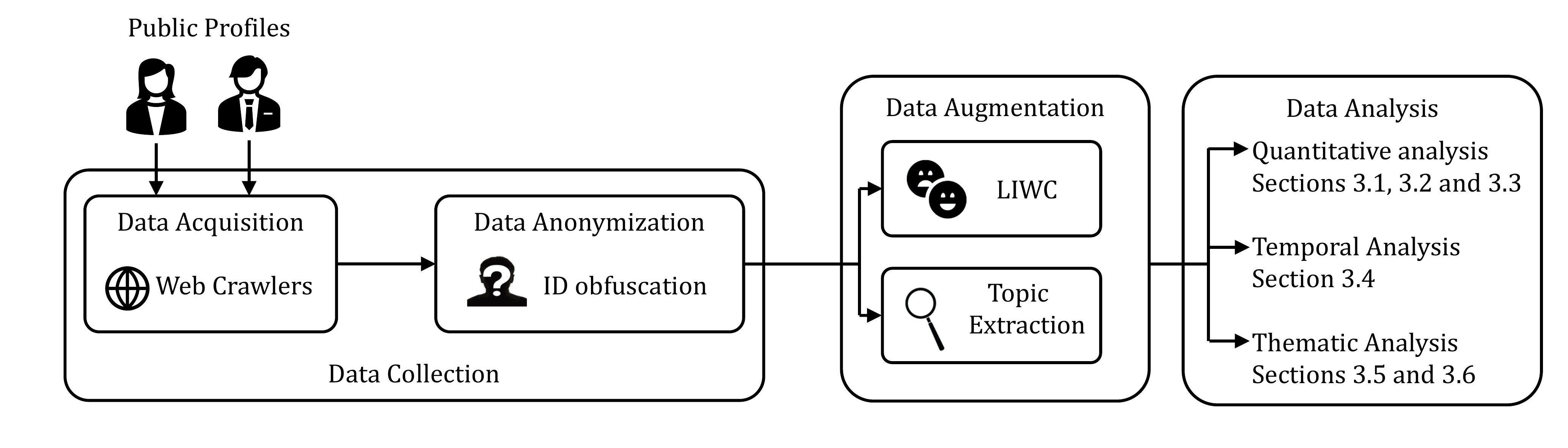}
    \caption{Data Collection and Analysis Methods.}
    \label{fig:methodology}
\end{figure}

In this section, we describe the methods we use to collect and process our dataset. The pipeline we follow is depicted in Figure~\ref{fig:methodology}. We first collect and anonymize the data through web crawlers. Then, we augment them with two different approaches, and, finally, we run our analyses.

\subsection{Data Acquisition}
\label{sec:dataset}

On Facebook and Instagram, a \emph{profile} can be followed by other profiles, i.e., its \emph{followers}. A profile with a large number of followers is also called an \emph{influencer}. Influencers post content (i.e., \emph{posts}), consisting of a photo, a video, or a plain text.\footnote{On Instagram, it is not allowed to create text-only posts.} The profile's followers, and anyone registered on the platform in the case of public profiles, can view, {\it like} and write \textit{comments} on the influencer's posts. In this paper, we are interested in monitoring the activity around the top Italian influencers. To this end, we build lists of the most popular influencers in Italy for different categories. First, we target Italian politicians, whose posts are known to reach a large audience and to produce peculiar interactions~\cite{Trevisan:2019}. We manually enumerate the Italian congressmen/congresswoman and political leaders, including in the list their official account on Instagram and Facebook, when present. Besides politicians, we build a list of \emph{General} influencers, composed of musicians, entertainers (i.e., actors/actresses and TV personalities), and athletes.\footnote{Some profiles refer to Music Bands or Sports Teams rather than physical persons.} In this case, we rely on public data to fill the list. For Instagram, we rely on HypeAuditor\footnote{\url{https://hypeauditor.com/}}, an online analytics platform, to get the list of top Italian influencers. For Facebook, we rely on the website \url{https://www.pubblicodelirio.it/} which offers an updated list of the most popular Italian influencers on Facebook, divided by category. Finally, among these profiles, we exclude those with minimal activity. The Facebook and Instagram lists mostly overlap, i.e., approximately $70 \%$ of individuals have an account on both platforms.

To build the dataset, we develop custom \textit{web crawlers} for both social networks. The crawlers, written in Python, use standard HTTP requests to collect data about the selected profiles. We run our crawlers continuously to download data at a slow speed, avoiding exceeding the social networks' websites' rate limits. For each monitored influencer, we download the profile metadata, e.g., the profile description and the number of followers, and all the generated \textit{posts}. For all the posts, our crawlers download the \textit{posting time}, i.e., when the post was written; the \textit{number of likes/reactions}\footnote{On Instagram, users \emph{like} posts, while, on Facebook, users \emph{react} to posts with a thumbs up or other five pre-defined emojis.}; and all the \textit{comments} written by \emph{any} profile in the first 24 hours after the posting time. For each comment, we collect information about the \textit{comment time} (with a minute granularity for Instagram and a day one for Facebook), the \textit{comment text}, the \textit{commenter identifier}, and the information whether the comment was a \textit{reply} to previous comment or to the original post. We store data on a Hadoop-based cluster and use Apache Spark and Python code for scalable processing.

\subsection{Data Anonymization}
\label{sec:privacy}

To respect the users' privacy, we take countermeasures to prevent users' re-identification. Firstly, we deployed our code following the Facebook and Instagram terms of usage described in their developers' terms and conditions.\footnote{\url{https://developers.facebook.com/terms/}} We anonymize the \textit{commenter identifier} as soon as it is read (before storing it to disk) using an irreversible hash function to remove any account identifier. This mechanism ensures the user is anonymized, respecting all the Data Use policies regarding the prohibited practices we could perform using the data. Moreover, the figures and results that we present only show aggregated results, preventing user-specific information disclosure.

To let other researchers replicate our results, we make our dataset available at~\cite{ourwebsite}. For this purpose, we remove from the public dataset all quasi-identifiers, i.e., information that can be used to indirectly re-identify a user. We do not share the content of the comments nor any other textual feature. Regarding the content of the comments, we provide summarized information regarding its psycholinguistic properties and possibly the discussed topic. As such, all \textit{Restricted Platform Data} that might be used to re-identify a user are removed.

\subsection{Data Augmentation}
\label{sec:dataset-methods-topic}

We run a twofold \textit{data augmentation} step (i)~to derive from each comment its \textit{psycholinguistic properties} and~(ii) to \textit{extract its topic}. 

\noindent
\textit{Psycholinguistic properties.}
We delve into the properties of comments using LIWC~\cite{Tausczik:2010}, a lexicon system that categorizes text into psycholinguistic properties. Words of the target language are organized as a hierarchy of categories and subcategories that form the set of LIWC attributes. Examples of attributes include linguistic properties (e.g., articles, nouns, and verbs), affect words (e.g., anxiety, anger, and sadness), and cognitive attributes (e.g., insight, certainty, and discrepancies). The hierarchy is customized for each language, with $83$ attributes for Italian. Notice that the LIWC methodology presents some limitations as, for instance, it cannot distinguish whether people are talking about themselves or referring to somebody else nor can detect sarcasm and analogies. In our analyses, we run LIWC on all the comments in our dataset and collect the obtained output, expressed as the percentage of words belonging to each attribute for each comment. For example, if in a comment 4 out of 20 words are related to a sad mood, the comment LIWC score of the ``Sadness'' attribute would be $20\%$.

\noindent
\textit{Topic extraction.} 
We manually choose the following topics related to the COVID-19 outbreak in Italy:
\begin{itemize}
    \item {\it COVID:}  general discussion about the virus and the pandemic.
    \item {\it Schools:} lectures were moved online on March 5$^{th}$.
    \item {\it Remote Working:} massively adopted for the first time as a consequence of the lockdown.
    \item {\it Dole:} the government issued support funds for unemployment during the lockdown.
    \item {\it Conspiracy:} comments related to conspiracy theories like the so-called \emph{Bill Gates conspiracy}~\cite{georgiou2020covid}, which claims that the former Microsoft CEO is planning to microchip individuals, the fear against 5G technology and the belief that it contributes to the pandemic and the existence of a new world order who caused the pandemic.
    \item {\it Home Cooking:} cooking and baking became very popular activities during the lockdown. In this regard, in Italy, the preparation of homemade bread, pizza, and in general bakery products caused a shortage in supply~\cite{BRACALE20201423,di2020eating}. 
\end{itemize}

To define our topic selection, we rely on the Google Trend platform, which analyzes the popularity of search queries in Google Search.\footnote{\url{https://trends.google.it/trends/explore?date=2020-3-11\%202020-5-18&geo=IT}} Looking at the top search keywords for the lockdown period: \textit{COVID} appears in different positions such as the 6$^{th}$ position with the term ``COVID-19''; \textit{School} appears in 8th position with ``Google Classroom"; \textit{Remote Working} appears in several positions with the name of popular video conference tools such as ``Zoom'' and ``Google Meet'' in the 3$^{rd}$ and 4$^{th}$, respectively; \textit{Dole} appears in the 2$^{nd}$ topic position with the term ``redundancy fund ''; \textit{Home Cooking} appears in the 12$^{th}$, 18$^{th}$, and 20$^{th}$ positions with ``Yeast'', ``Bread'', and ``Dough'' respectively. Finally, while the \textit{Conspiracy} is not in the Google Trends trending topics, we select it as a case study for the spread of fake news in social networks, and its importance has been highlighted in the recent literature~\cite{5gConspiracy,bruns2020covid19,celestini2020information}. 

For each topic, we manually select the related key Italian terms. For example, for the topic \emph{COVID-19}, we look for words like COVID, pandemic, and coronavirus in the Italian language. We use stemming to reduce inflected words. We look for the topic terms in the comment corpus, and, whenever we find a match, we flag the comment as discussing the topic. Notice that since terms of different topics may be found in the corpus of the same comment, we can flag a comment as discussing multiple topics. In Section~\ref{sec:4.6}, we manually validate the accuracy of our methodology for topic detection showing that the False Positive Rate is below 20\% in all cases.

\subsection{Data Analysis}

To extract knowledge from the data, we perform our analyses using the following methods: 

\noindent \textit{Quantitative data analysis.} We measure users' and influencers' interactions before, during, and after the lockdown. We describe the results in Sections~\ref{sec:charact},~\ref{sec:4.1} and~\ref{sec:4.3}.

\noindent \textit{Temporal analysis.} We study how temporal patterns in users' interactions varied and show results in Section~\ref{sec:4.4}.

\noindent \textit{Thematic Analysis.} We study changes in the psycholinguistic properties of the comments (Section~\ref{sec:4.5}) and quantify the spread of COVID-related topics (Section~\ref{sec:4.6}).

\section{Results}
\label{sec:results}

\subsection{The dataset at a glance}
\label{sec:charact}

\begin{table}[t]
    \renewcommand{\arraystretch}{1.25}
    \footnotesize
    \caption{Dataset summary}
    \label{tab:dataset}
    \centering
    \begin{tabular}{lrrrr}
    \hline & &     Profiles & Posts    & Comments \\
    \hline
    \hline
    & \textit{\textbf{Instagram}} &       $284$ &   $44\,918$   & $17.6$ M \\
    \hline
    & Politicians &    $109$ &   $27\,877$   & $6.9$ M \\    \cdashline{1-2}
    \multirow{3}{*}{\rotatebox[origin=2]{90}{\scriptsize \textit{General}}}
    & Musicians &       $54$  &   $5\,172$    & $4.3$ M \\
    & Entertainers &        $35$  &   $3\,876$    & $2.6$ M \\
    & Athletes &       $86$  &   $7\,993$    & $3.6$ M \\
    \hline
    \hline
    & \textit{\textbf{Facebook}} &       $355$ &   $95\,420$   & $37.2$ M \\
    \hline
    & Politicians &    $232$ &   $78\,697$   & $31.8$ M \\ \cdashline{1-2}
    \multirow{3}{*}{\rotatebox[origin=2]{90}{\scriptsize \textit{General}}}
    & Musicians &       $51$  &   $4\,643$    & $2.3$ M \\
    & Entertainers &        $22$  &   $10\,838$   & $2.8$ M \\
    & Athletes &       $51$  &   $1\,242$    & $0.6$ M \\
    \hline
    \end{tabular}
\end{table}

We summarize our dataset in Table~\ref{tab:dataset}, which reports various metrics separately by the social network and profile category. Overall, our dataset spans the first six months of 2020. Figure~\ref{fig:events} reports the timeline of the main events related to the COVID-19 pandemic in Italy. We define the following three periods in 2020: \textit{pre-lockdown} (before lockdown) period from January 1$^{st}$ to February 19$^{th}$, hence before the first Italian case of COVID-19; \textit{lockdown} period from March 11$^{th}$ to May 18$^{th}$, where people were forced to stay home; and \textit{post-lockdown} (after lockdown) from May 18$^{th}$ to July 1$^{st}$. In the following, we use the terms pre-lockdown and before lockdown interchangeably, as well as post-lockdown and after lockdown.  Notice that when we compare the pre-lockdown and lockdown periods, we neglect the data between February 19$^{th}$ and March 11$^{th}$ as it represents a transient period with increasing restrictions.

\begin{figure}
\begin{center}
    \begin{subfigure}{0.48\columnwidth}
        \includegraphics[width=\columnwidth]{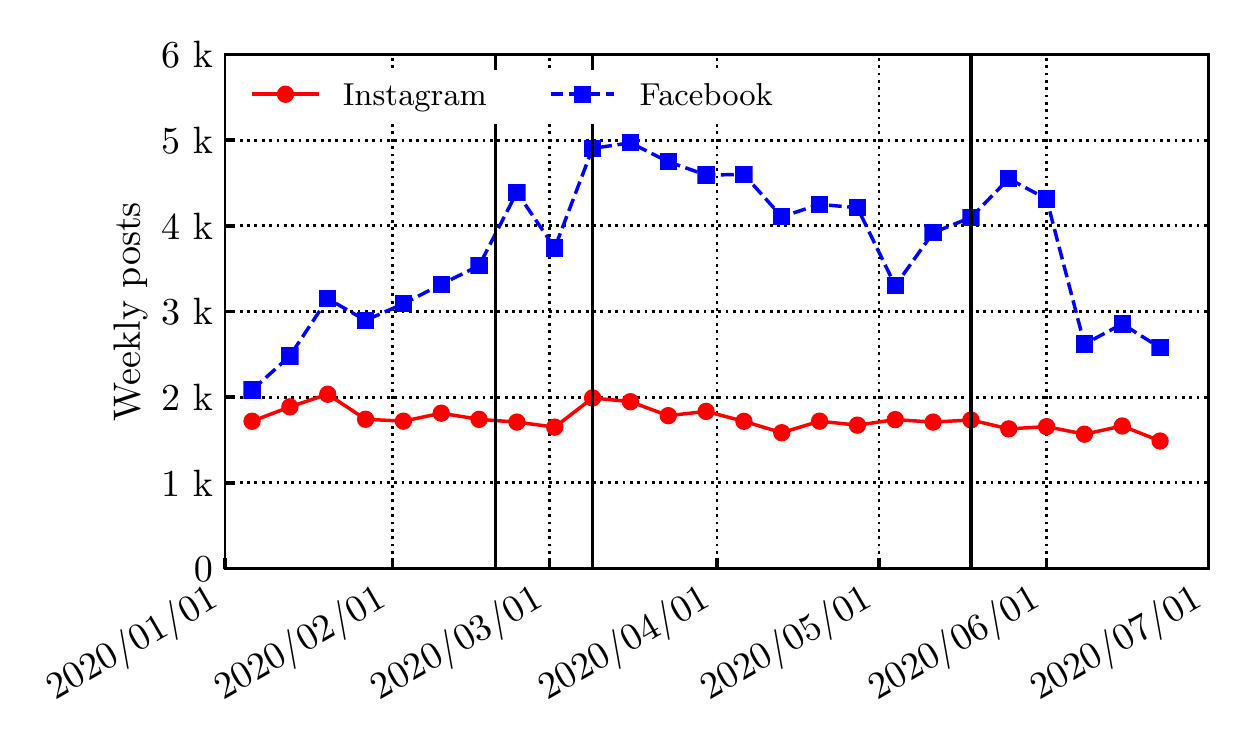}
        \caption{Posts.}
        \label{fig:post_volume}
    \end{subfigure}
    \begin{subfigure}{0.48\columnwidth}
        \includegraphics[width=\columnwidth]{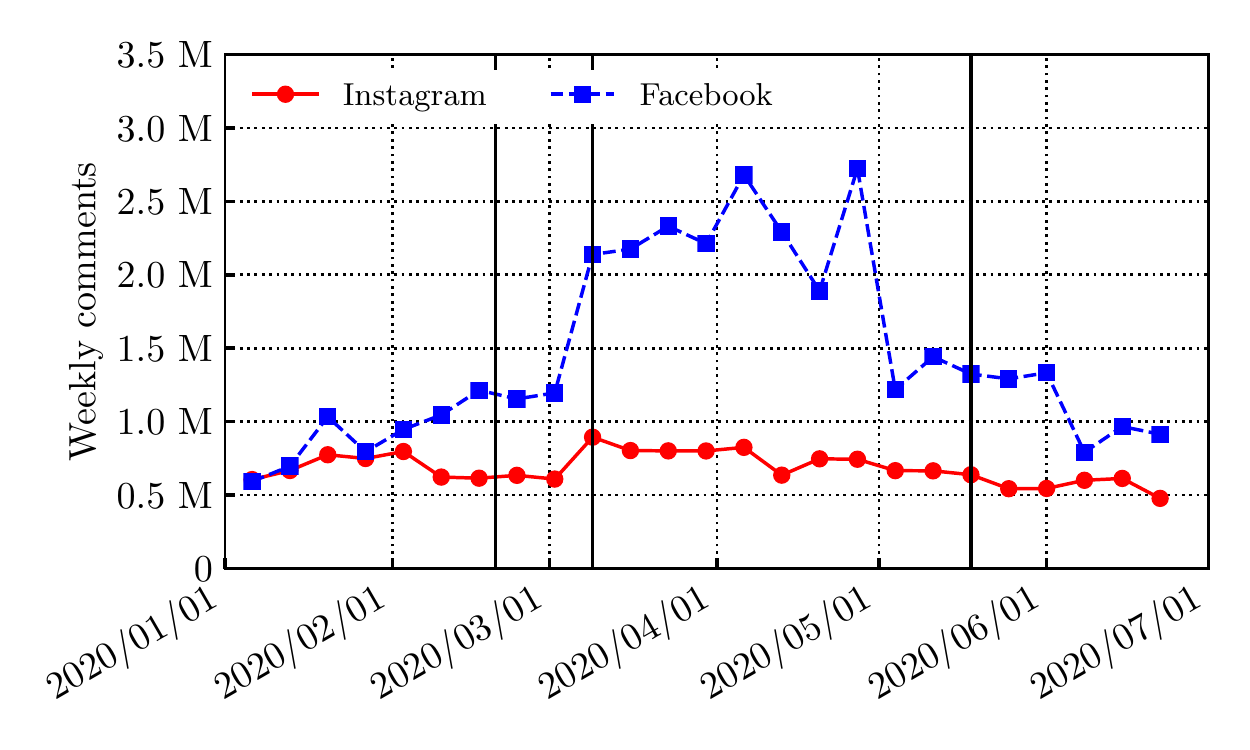}
        \caption{Comments.}
        \label{fig:comment_volume}
    \end{subfigure}
    \begin{subfigure}{0.48\columnwidth}
        \includegraphics[width=\columnwidth]{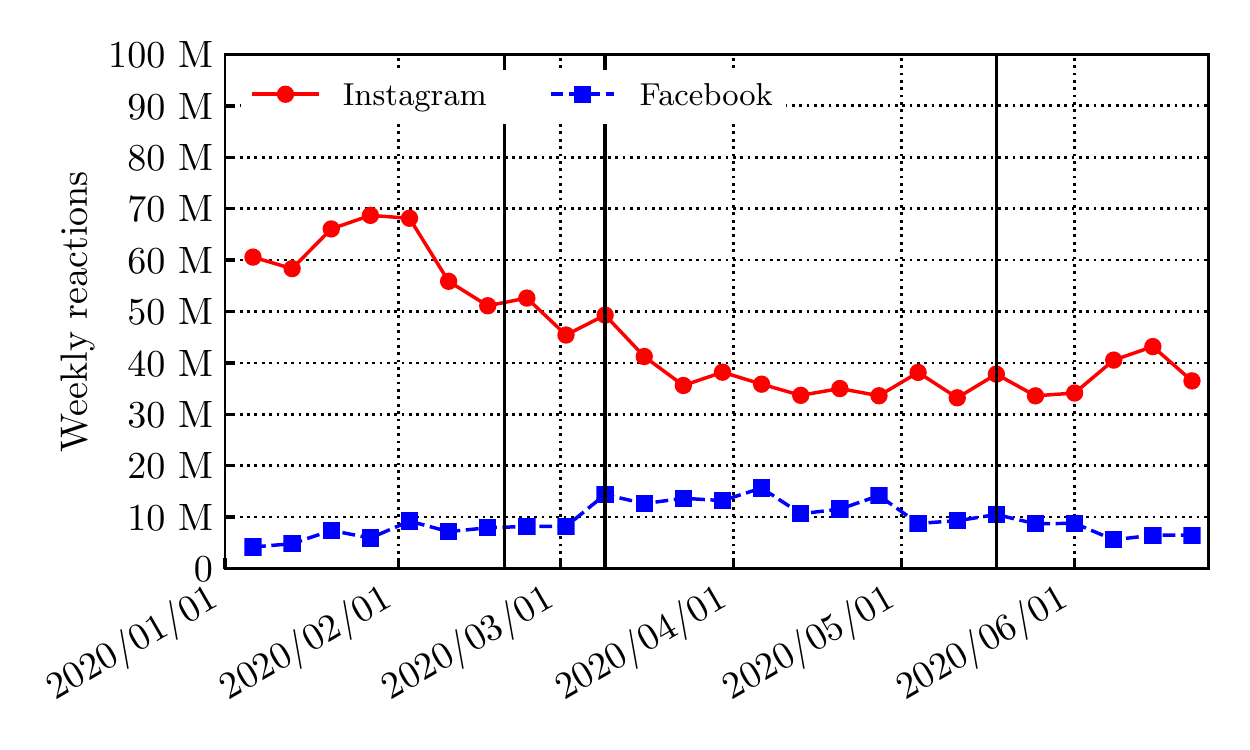}
        \caption{Reactions.}
        \label{fig:reaction_volume}
    \end{subfigure}
    \caption{Number of weekly posts published, reactions and comments received, per Instagram  and Facebook.}
    \label{fig:all_volume}
    \end{center}
\end{figure}

In total, we monitored $639$ influencers which published $140\,$ thousands of posts. On Instagram, we monitored fewer Politicians than on Facebook as some of them are not popular and obtain a small number of interactions (likes or comments). In total, the posts received approximately $55$ million comments from $35$ million distinct profiles (or \emph{commenters}), with Politicians' posts attracting most of the interactions on both social networks.  In Figure~\ref{fig:all_volume}, we provide an overview of the temporal evolution of our dataset, separately for Instagram and Facebook. Figure~\ref{fig:post_volume} depicts the number of posts created by the selected influencers, week by week. The solid red line refers to Instagram, while the dashed blue line to Facebook. The three black vertical lines represent the events reported in Figure~\ref{fig:events}. The two social networks exhibit different trends. Facebook shows a $47$\% increase in the number of posts during the lockdown weeks, while Instagram presents almost a flat trend with  $-3$\% of posts in the whole period. Considering the influencers' category, we observe that most of the posts are generated by Politicians.\footnote{In the Appendix, we show the distribution of the number of daily posts per profile.}

Similar considerations hold for the volume of comments received by the posts (Figure~\ref{fig:comment_volume}). Facebook comments increased by $125$\% during the lockdown, while Instagram comments increased only by $+8$\%. Interestingly, the number of likes/reactions received by the posts (Figure~\ref{fig:reaction_volume}) exhibit different trends. On Instagram, the number of likes/reactions received by posts shows a substantial decrease during the lockdown period ($-39$\%), while on Facebook, it increased by $+83$\%. Considering the absolute numbers, however, Instagram posts received a greater number of likes than Facebook. This shows different usage patterns between the two social networks.

\subsection{Quantification of the interactions}
\label{sec:4.1}

\begin{figure}
    \begin{subfigure}{0.48\columnwidth}
        \includegraphics[width=\columnwidth]{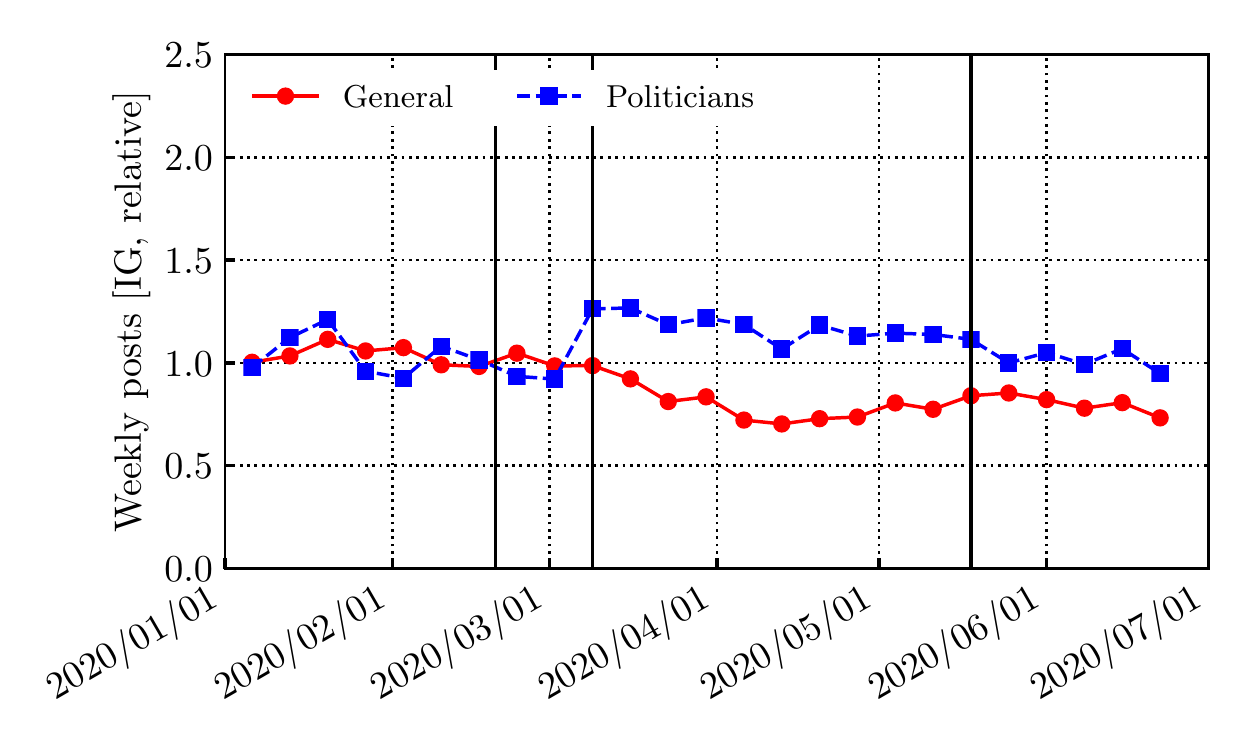}
        \caption{Instagram.}
        \label{fig:post_category_ig}
    \end{subfigure}
    \begin{subfigure}{0.48\columnwidth}
        \includegraphics[width=\columnwidth]{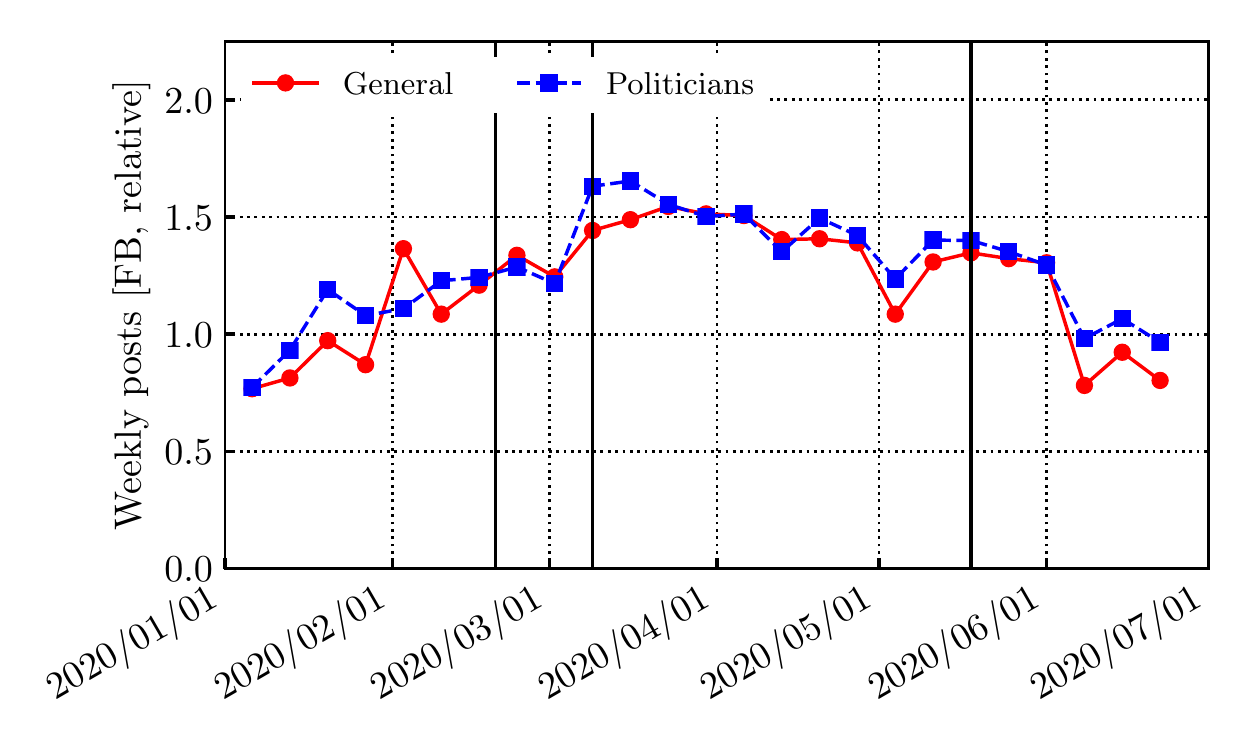}
        \caption{Facebook.}
        \label{fig:post_category_fb}
    \end{subfigure}
    \caption{Number of posts, normalized over the pre-lockdown period.}
    \label{fig:post_category}
\end{figure}

We now provide quantitative figures to describe the influencers posting behaviour and the commenters' interactions. We start from what emerged in the previous subsection, which already showed an impact of the COVID-19 on the users' behavior on the two platforms. In Figure~\ref{fig:post_category}, we report the weekly number of posts created by the monitored influencers, separately for profile category (Politicians or General) and social network. To ease the visualization, here we normalize the values by the average of the pre-lockdown period so that variations are more evident (aggregate absolute values are presented in Figure~\ref{fig:all_volume}).  

Figure~\ref{fig:post_category_ig} shows the number of posts for Instagram, in which the dashed blue line refers to the Politicians, while the red solid lines group together the \emph{General} category. In this case, we observe a different behavior between the two groups: the number of posts for Politicians increased by $12$\% during the lockdown, while, for General, it decreased by $22$\%. Differently, for Facebook (Figure~\ref{fig:post_category_fb}), both groups show an increasing number of posts by $+38$\% and $+36$\% respectively for General and Politicians. Interestingly, regardless of the magnitude of the increase during the lockdown, in the post-lockdown both social networks show a decreasing trend with the number of posts returning to the same level as the pre-lockdown period. Considering the absolute number of posts per influencer (reported in the Appendix), Politicians produced more posts than the other categories and even increased their rate during the lockdown.

\begin{figure}
    \begin{subfigure}{0.48\columnwidth}
        \includegraphics[width=\columnwidth]{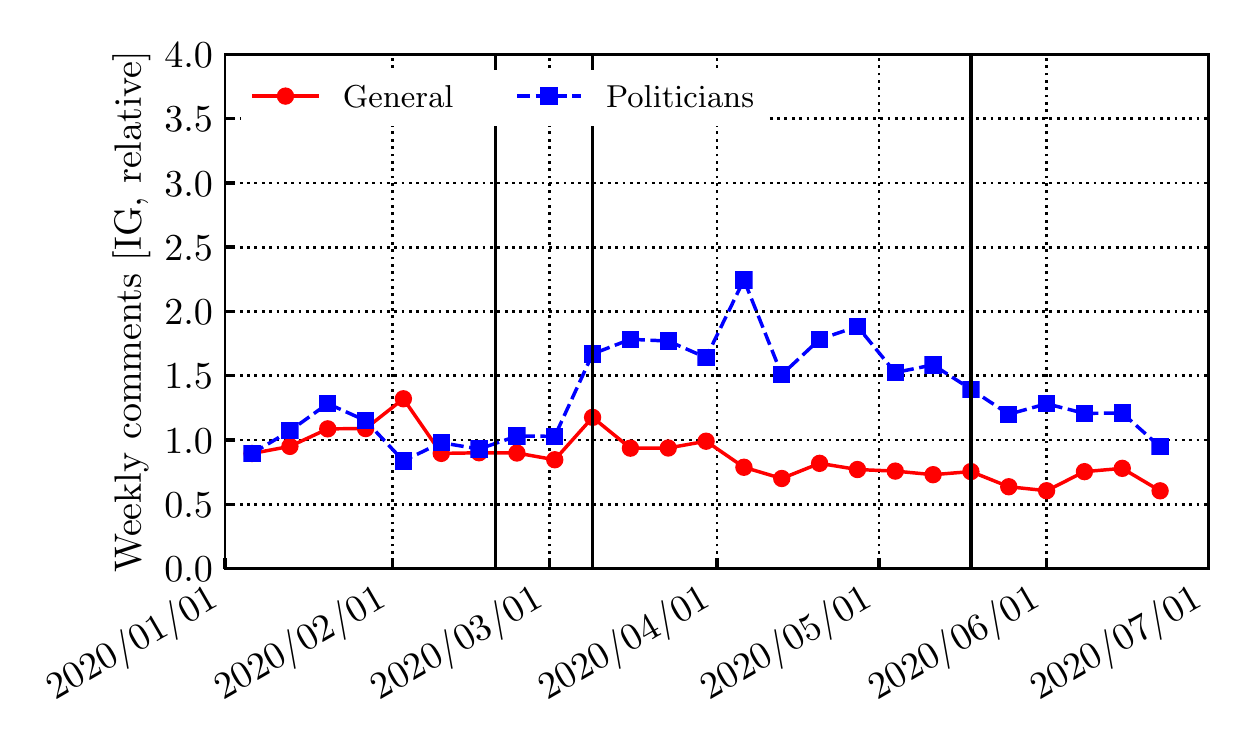}
        \caption{Instagram.}
        \label{fig:comments_category_ig}
    \end{subfigure}
    \begin{subfigure}{0.48\columnwidth}
        \includegraphics[width=\columnwidth]{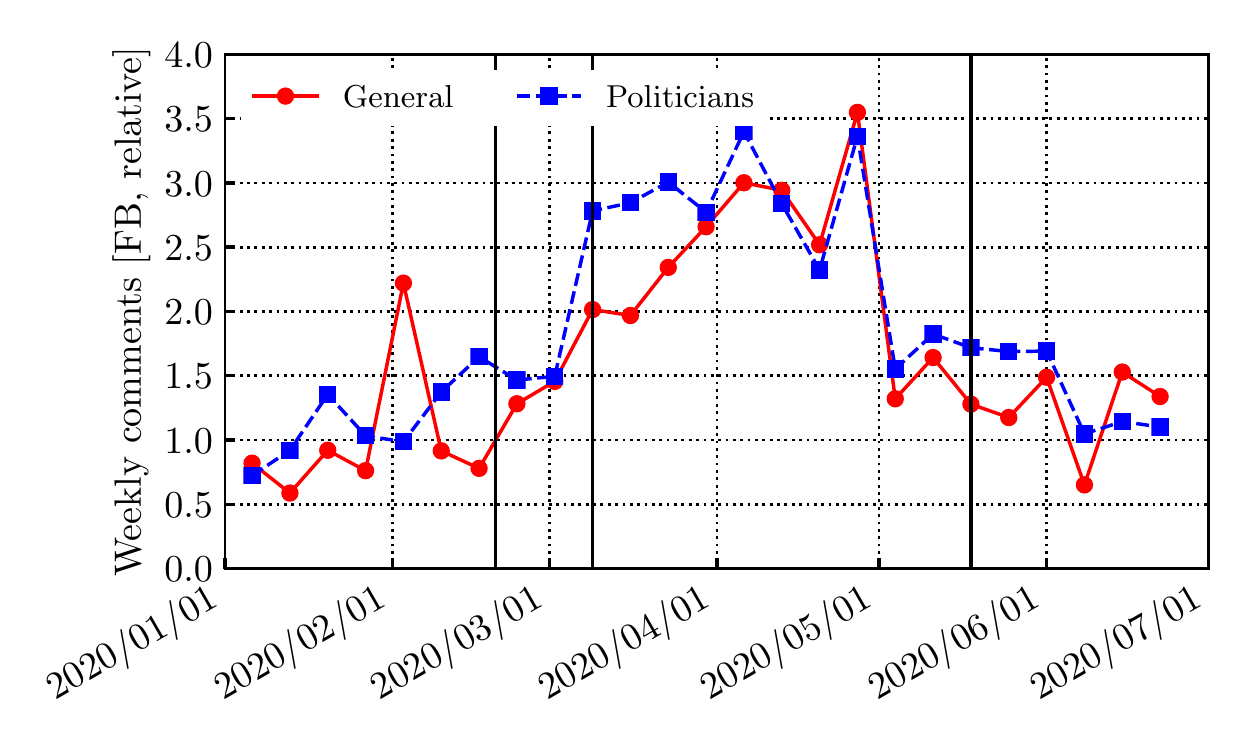}
        \caption{Facebook.}
        \label{fig:comments_category_fb}
    \end{subfigure}
    \caption{Number of comments, normalized over the pre-lockdown period.}
    \label{fig:comments_category}
\end{figure}

The differences between the social networks are more evident if we look at the comments these posts received. Figure~\ref{fig:comments_category} shows the weekly volume of comments normalized by the average in the pre-lockdown period. In the first week of lockdown, the comments suddenly increased on Politicians, with $+70$\% and $+280$\% for Instagram and Facebook, respectively. For Politicians' posts, the volume of comments kept its high value for the entire lockdown period. We notice that short posts (with less than $50$ characters) received, in median, half of the comments than long ones -- see the Appendix for the complete figures. Considering General profiles, they showed an increase in volume during the lockdown only on Facebook ($+120\%$), even if not as pronounced as for Politicians ($+130\%$). As an anecdote, notice the peak of General profiles on Facebook on the first week of February (Figure~\ref{fig:comments_category_fb}). It corresponds to the 2020 \emph{Festival di Sanremo}, which is the most popular Italian music competition. Indeed, we observe a considerable increase of likes/reactions in the Musicians' profiles.

\begin{figure}
    \begin{subfigure}{0.48\columnwidth}
        \includegraphics[width=\columnwidth]{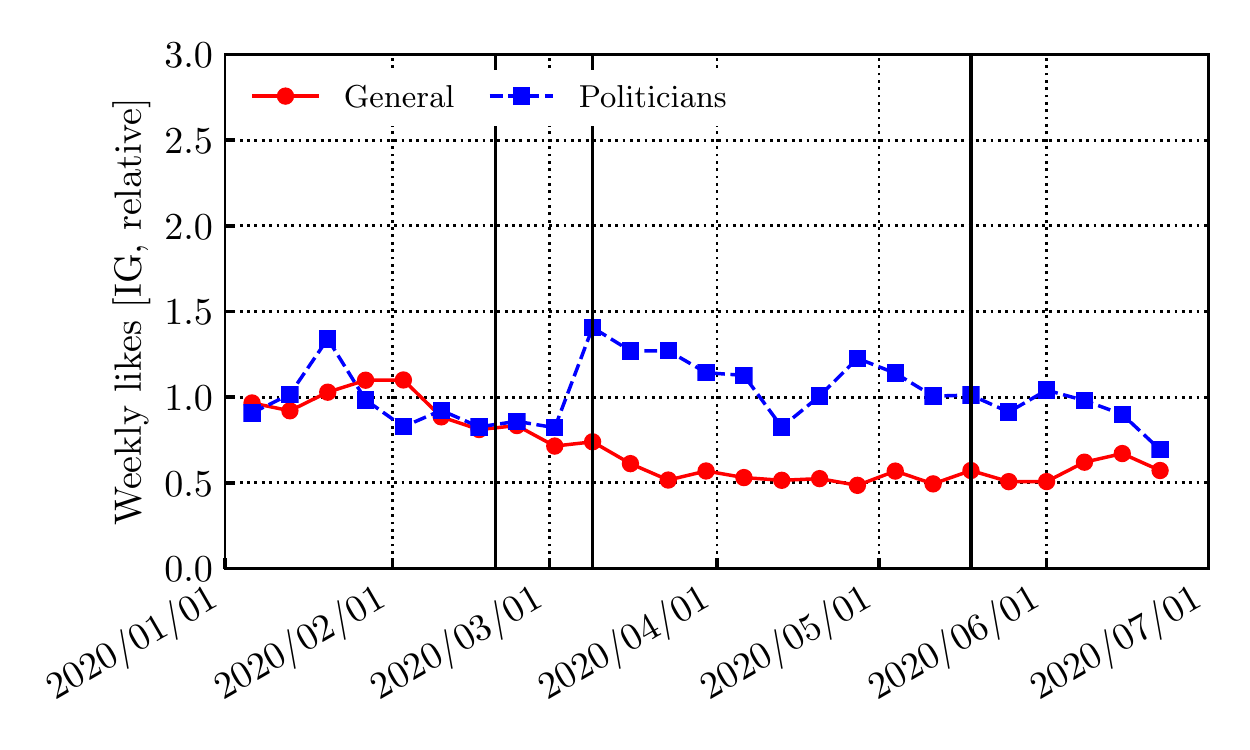}
        \caption{Instagram.}
        \label{fig:like_category_ig}
    \end{subfigure}
    \begin{subfigure}{0.48\columnwidth}
        \includegraphics[width=\columnwidth]{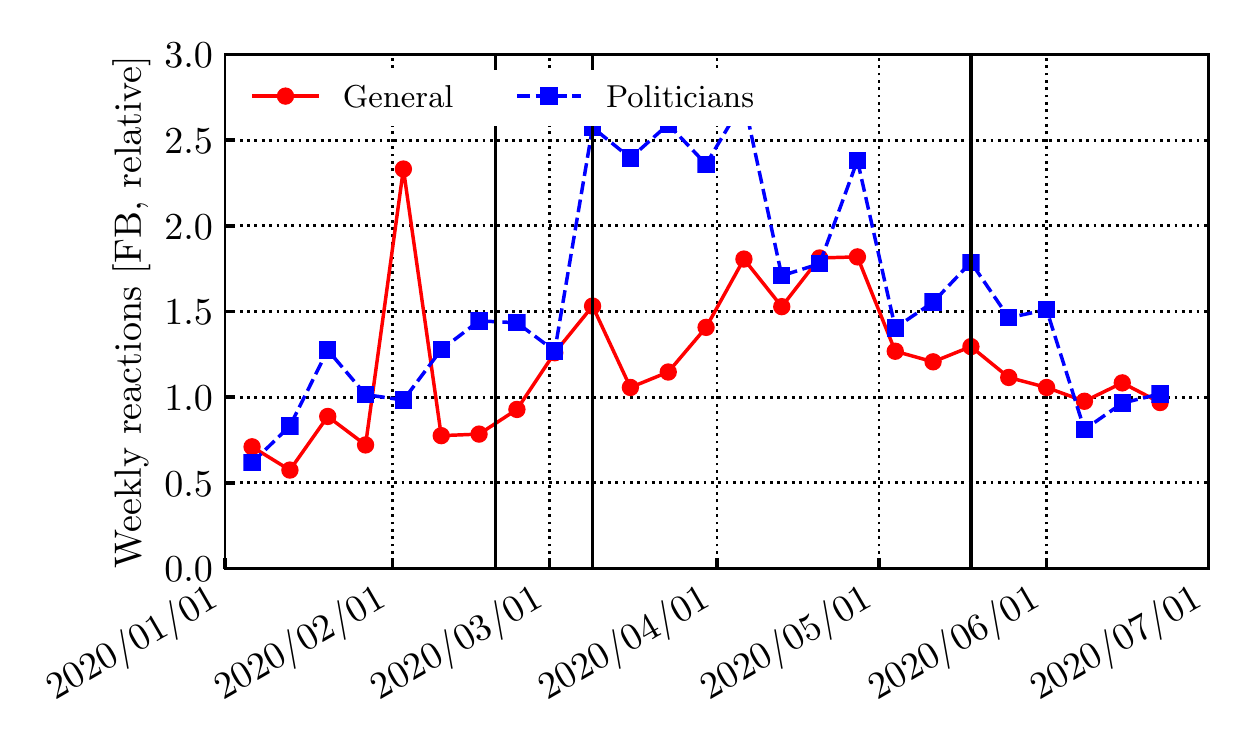}
        \caption{Facebook.}
        \label{fig:like_category_fb}
    \end{subfigure}
    \caption{Number of likes/reactions, normalized over the pre-lockdown period.}
    \label{fig:like_category}
\end{figure}

In Figure~\ref{fig:like_category}, we consider the number of likes (in the case of Instagram) and reactions (in the case of Facebook) the monitored posts received. Likes/reactions to Politicians' posts increased in both social networks by $+98$\% and $+15$\% for Facebook and Instagram, respectively. However, for Instagram, the General profiles obtained $-43$\% likes. Recalling that the volume of posts on Instagram decreased only by $22$\%, the higher reduction in likes suggests that, other than the smaller number of posts, the decrease is also driven by a users' disinterest in the influencers belonging to the general category. Indeed, the drop in likes and comments is generalized to all the three sub-categories of General, especially for the Athletes. The same considerations hold for the comments.

\begin{figure}
    \centering
    \includegraphics[width=0.7\columnwidth]{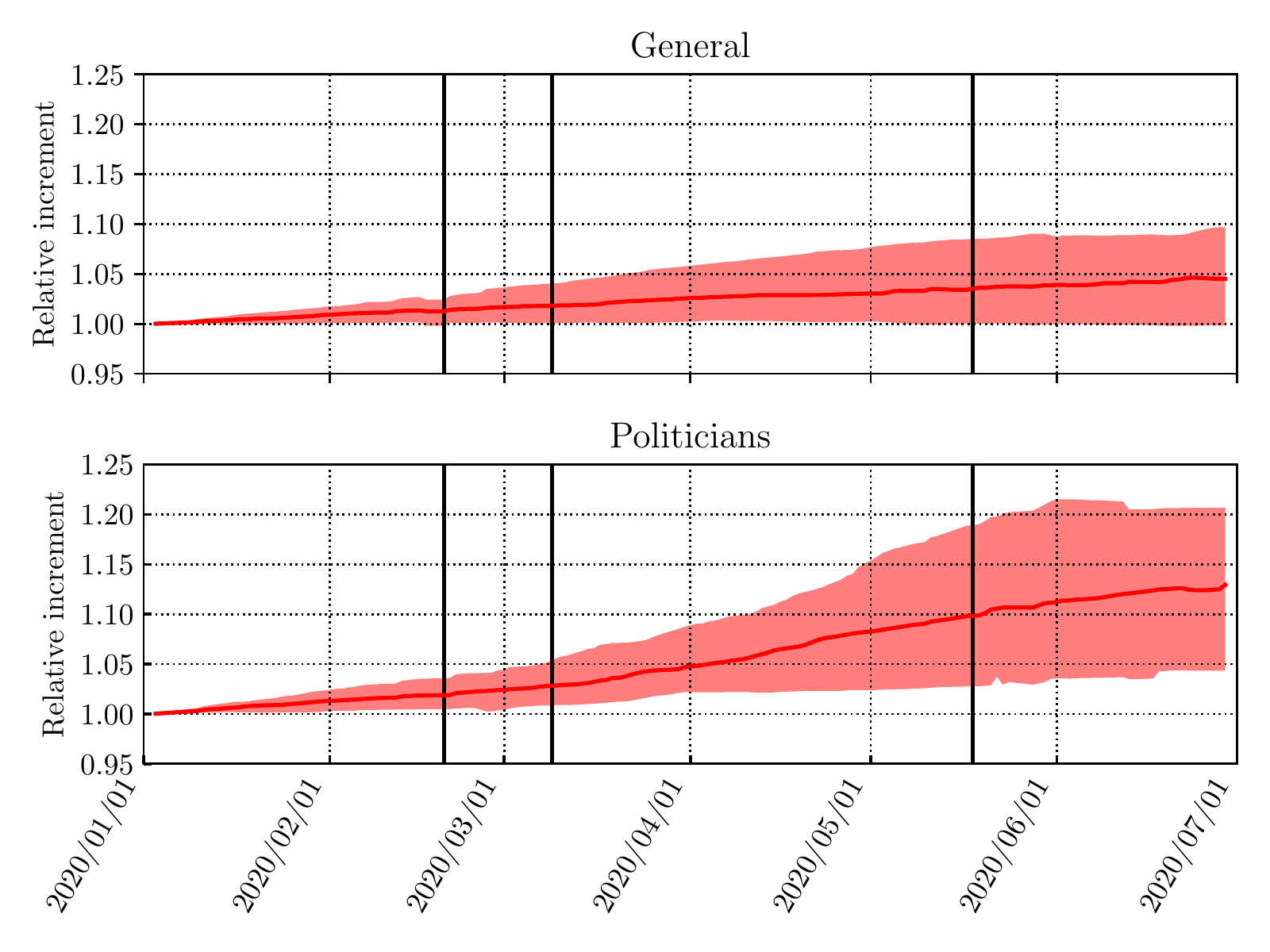}
    \caption{Relative increase of followers in the first six months of 2020 (Instagram). 25$^{th}$, 50$^{th}$ and 75$^{th}$ percentiles are shown. }
    \label{fig:follower}
\end{figure}

We now study whether the monitored Italian influencers gained popularity in terms of the number of followers. We restrict this analysis to Instagram only, as for Facebook we could not collect historical metadata about influencers. For each influencer, we compute the relative variation of followers with respect to the first day of the dataset, i.e., January 1$^{st}$, 2020. For example, if an influencer had $1\,000$ followers on January 1$^{st}$ and $1\,500$ on May 15$^{th}$, the relative variation on May 15$^{th}$ is~$1.5$. We show the results in Figure~\ref{fig:follower}, separately for Politicians and General profiles. The red line shows the median variation, i.e., the median value over all profiles, while the area spans from the first to the third quartile. As expected, the variation is usually positive as rarely a user \emph{unfollows} a profile. We observe changes in the growth rate, especially for Politicians. To quantify the rate at which the number of followers increased, we perform a linear regression over the median variation. First, we analyze General profiles. Before the lockdown, in median, they weekly increased the number of followers by $0.22\%$. During the lockdown, this number slightly increased to $0.25\%$ weekly. Analyzing the Politicians, they were already acquiring more followers before the lockdown, with a median weekly increase of $0.66\%$. During the lockdown, this number raised to $0.98\%$ weekly. In the entire first six months of 2020, for Politicians, the median growth of followers reached $13\%$, while for General influencers $5\%$ only. This difference might be due to the importance and the impact of political choices during the lockdown. We discuss in-depth the implications and potential causes in Section \ref{sec:discusion}. Considering the General category profiles, we notice that Athletes are those with the lowest increase, potentially linked to the forced interruption of all sporting activities. In the Appendix, we extend this analysis to compare this growth with the one observed in the second semester of 2019 and July-August 2020.

\subsection{Level of debate}
\label{sec:4.3}

\begin{figure}[t]
    \begin{subfigure}{0.48\columnwidth}
        \includegraphics[width=\columnwidth]{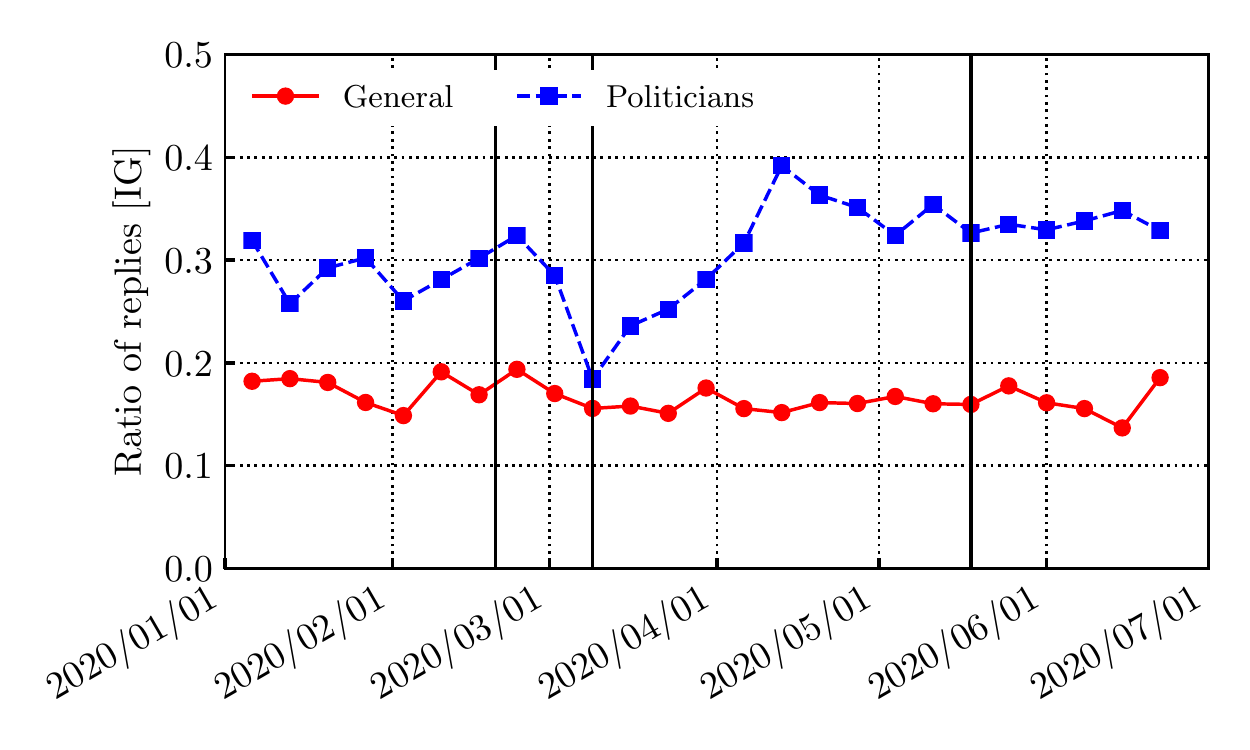}
        \caption{Instagram.}
        \label{fig:reply_share_ig}
    \end{subfigure}
    \begin{subfigure}{0.48\columnwidth}
        \includegraphics[width=\columnwidth]{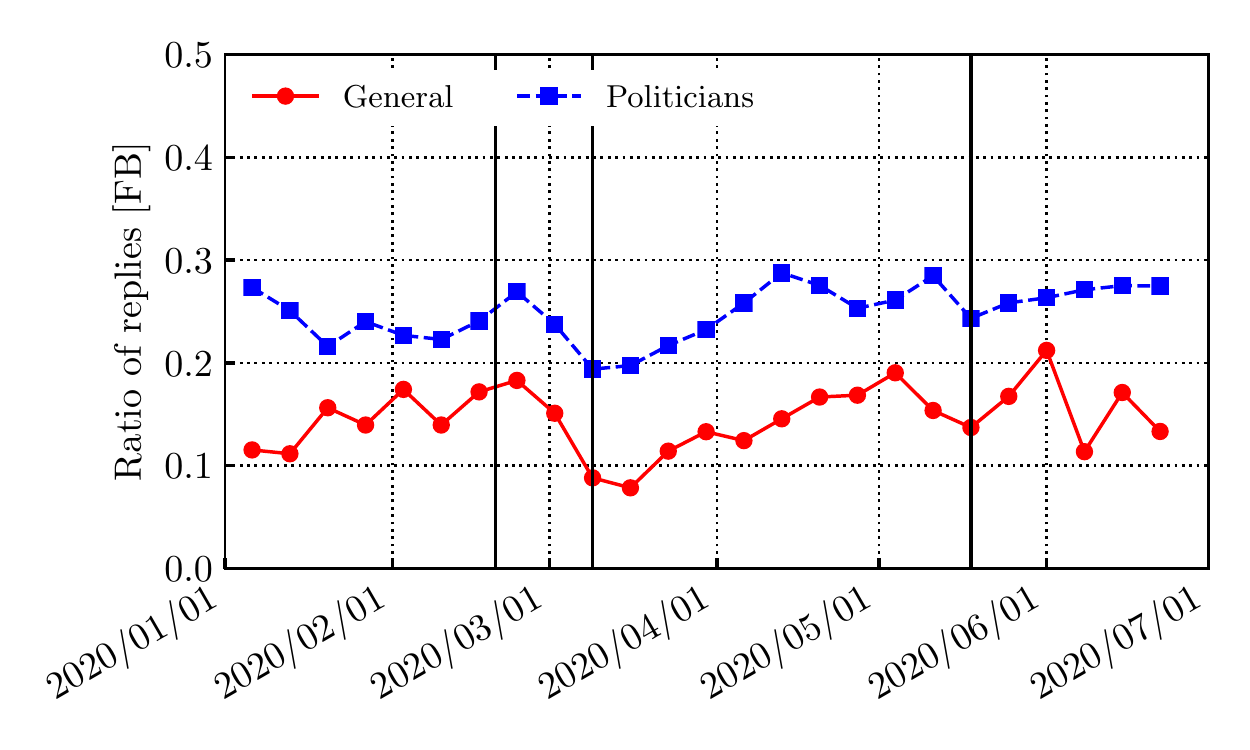}
        \caption{Facebook.}
        \label{fig:reply_share_fb}
    \end{subfigure}
    \caption{Level of debate: ratio of comments which are replies to other comments.}
    \label{fig:reply_share}
\end{figure}

We estimate the level of debate around posts as the fraction of comments that are \emph{replies} to other comments. A user can comment on a post replying directly to another user's comment on both social networks.  This allows us to evaluate whether people engage in conversations since replies indicate that users interact with each other with questions, answers, arguments, etc.

We plot the ratio of replies over all comments week by week in Figure~\ref{fig:reply_share}, separately by social network and profile category. Focusing on Instagram (Figure~\ref{fig:reply_share_ig}), we observe that, on average, before the lockdown, approximately $30\%$ of comments on Politicians' posts (blue dashed line) are replies. This percentage decreases to less than $15-20\%$ for the General profiles (solid red line). This confirms the findings of our previous work~\cite{Trevisan:2019}, where we showed how Politicians receive more comments per follower, with a larger share of replies ($+55$\%) than the other categories. However, during the first weeks of lockdown, we notice a drastic decrease of replies for Politicians' profiles, with only $18 \%$ of replies. They return gradually to pre-lockdown values in the fourth week and finally settle on $35\%$ on the last weeks of lockdown. Considering the post-lockdown we can see how users tend to reply more compared to the pre-lockdown period. This means that, at the beginning of the lockdown, the debate around Politicians decreased, the number of isolated comments to posts increased, and users were less engaged relatively less in discussion with each other. However, during the second part of the lockdown and in the weeks after, the debate has come back even higher. Looking at the comments on the General profiles' posts, on the contrary, we observe little variation with around $13 \%$ replies. This suggests that the debate around these influencers did not change as much as for the Politicians. 

Focusing on Facebook (Figure~\ref{fig:reply_share_fb}), similar considerations hold. The debate level decreased at the beginning of the lockdown and reached values (sightly) higher than before in the post-lockdown. Interestingly, this holds for both profile categories, confirming that, on Facebook, the COVID-19 has an effect on most of the profiles. We comment on these results, conjecturing possible causes in Section~\ref{sec:discusion}.

Finally, we investigate the comment length as it could indicate the users' interest in arguing their claims. During and after the lockdown, comments are slightly longer. However, the distributions have limited discrepancy. Hence they do not provide any statistical significance.

\subsection{Daily and weekly patterns}
\label{sec:4.4}

We now study the activity patterns to understand changes in the users' habits during the COVID-19 outbreak. We focus on the daily and weekly patterns and restrict our analysis only on Instagram, as, for Facebook, the comment creation time is available only with day granularity.   

\begin{figure}[t]
    \centering
    \includegraphics[width=0.6\columnwidth]{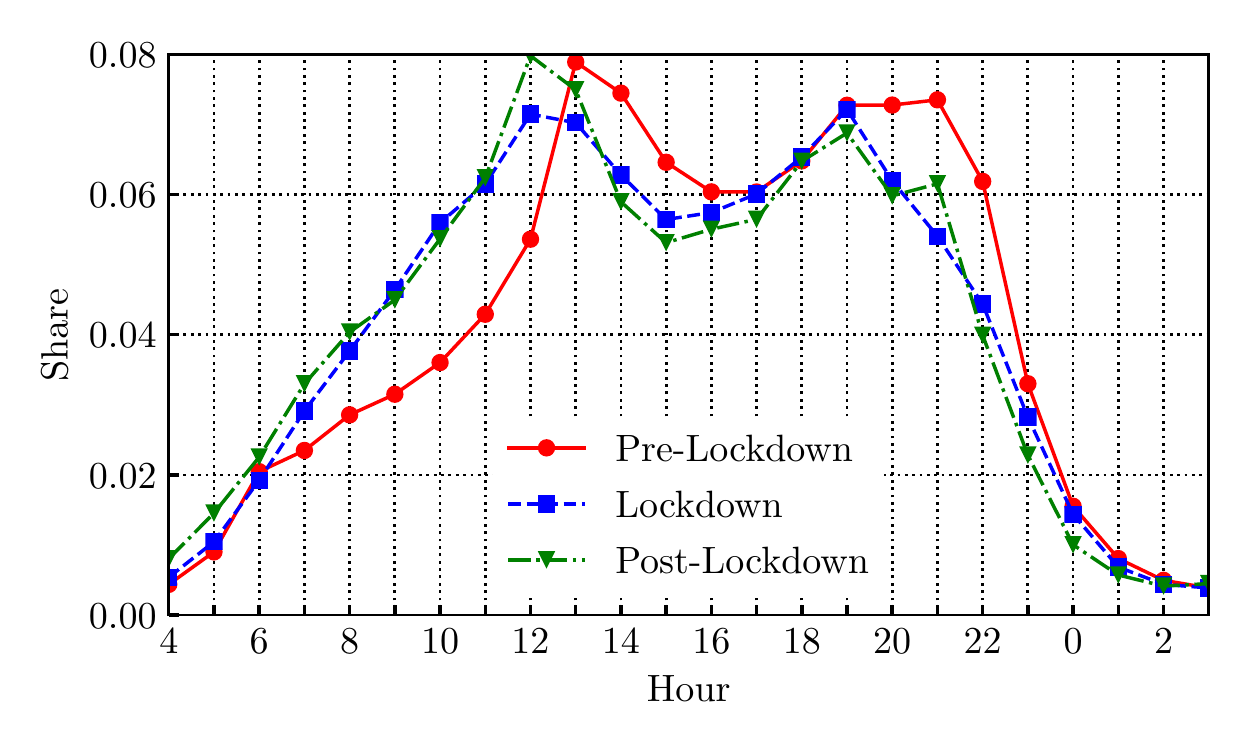}
    \caption{Distribution of Instagram comments over the day for different periods. Notice that the $x$-axis begins at 4 AM.}
    \label{fig:hour}
\end{figure}

We start our analysis by focusing on the daily pattern. Here, we make no distinction between Politicians and General and provide the overall picture only. In Figure~\ref{fig:hour}, we show the distribution of the comment creation time across the hours of the day. The $24$ hours are arranged on the $x$-axis, starting at the time at which the comment rate is lower ($4$ AM). The $y$-axis represents the share of comments created within the given hour, and different lines correspond to the different periods. It is no surprise that the patterns present two peaks roughly at lunch and dinner time. However, the pre-lockdown curve (solid red) presents considerable differences with respect to the other lines.  Indeed, during and after the lockdown (blue and green dashed lines, respectively), we observe a $36$\% increase in the morning activity from $7$ AM to $12$ PM. The activity is instead lower (in proportion) during the first hours of the afternoon and in the evening. Interestingly, the lockdown and post-lockdown curves are relatively similar (differences are limited in $\pm 11$\%), meaning that the variation of habits persists after the end of the lockdown. We will discuss possible causes and implications of this long-term shift in behavior in Section \ref{sec:discusion}.

\begin{figure}[t]
    \centering
    \includegraphics[width=0.7\columnwidth]{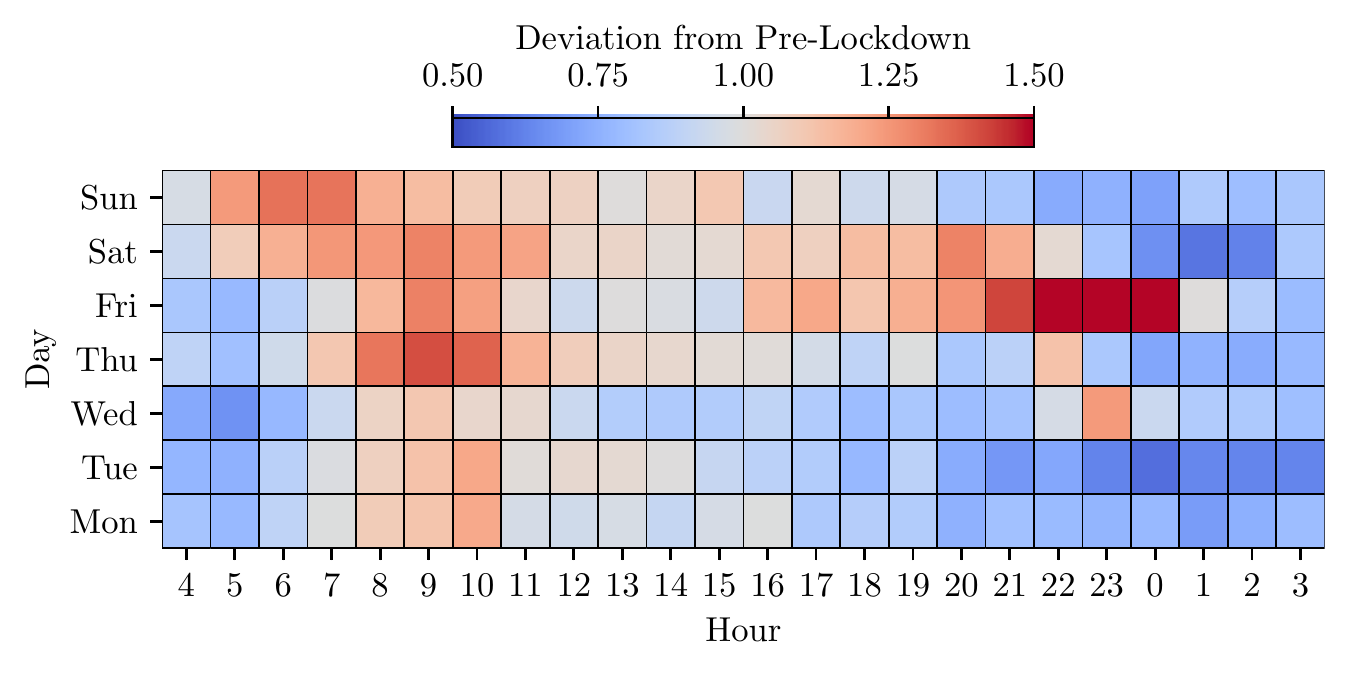}
    \caption{Variation of the weekly pattern of Instagram comments during lockdown compared to the pre-lockdown period. Notice that the $x$-axis begins at 4 AM.}
    \label{fig:week}    
\end{figure}

We now focus on the weekly patterns, breaking down users' activity over the seven days of the week. Rather than providing separate figures for different periods, we opt to show only the lockdown variations compared to the pre-lockdown period, offering a more concise view. In Figure~\ref{fig:week}, we arrange the weekdays on different rows, while columns represent the $24$ hours of the day. For both the lockdown and pre-lockdown periods, we compute for each pair $(day,hour)$ the share of comments it contains over the total number of comments (for that period). Then, in the figure, each cell indicates the deviation (increase or decrease) with different colors during the lockdown compared to the pre-lockdown, computed as the ratio between the two shares. We first observe that the increase in the morning activity, already shown in Figure~\ref{fig:hour}, is present similarly over all the weekdays. Moreover, we notice a consistent increase (up to $50$\%) in the activity on Friday and Saturday afternoons and evenings. This increase is likely linked to the prohibition of social life during the lockdown. The figure also shows increased activity during the early morning hours ($5$ AM - $6$ AM) of Saturday and Sunday. This deviation gives evidence on how the usage pattern varied, with smaller differences during working days than weekends.

Once the lockdown ends, we observe that the activity on Friday and Saturday evenings returns to the pre-lockdown levels, but the higher volume in the morning persists. We report the corresponding figure in the Appendix.

\subsection{Psycholinguistic properties of comments}
\label{sec:4.5}

\begin{figure}[!t]
    \begin{subfigure}{\columnwidth}
        \centering
        \includegraphics[width=0.7\columnwidth]{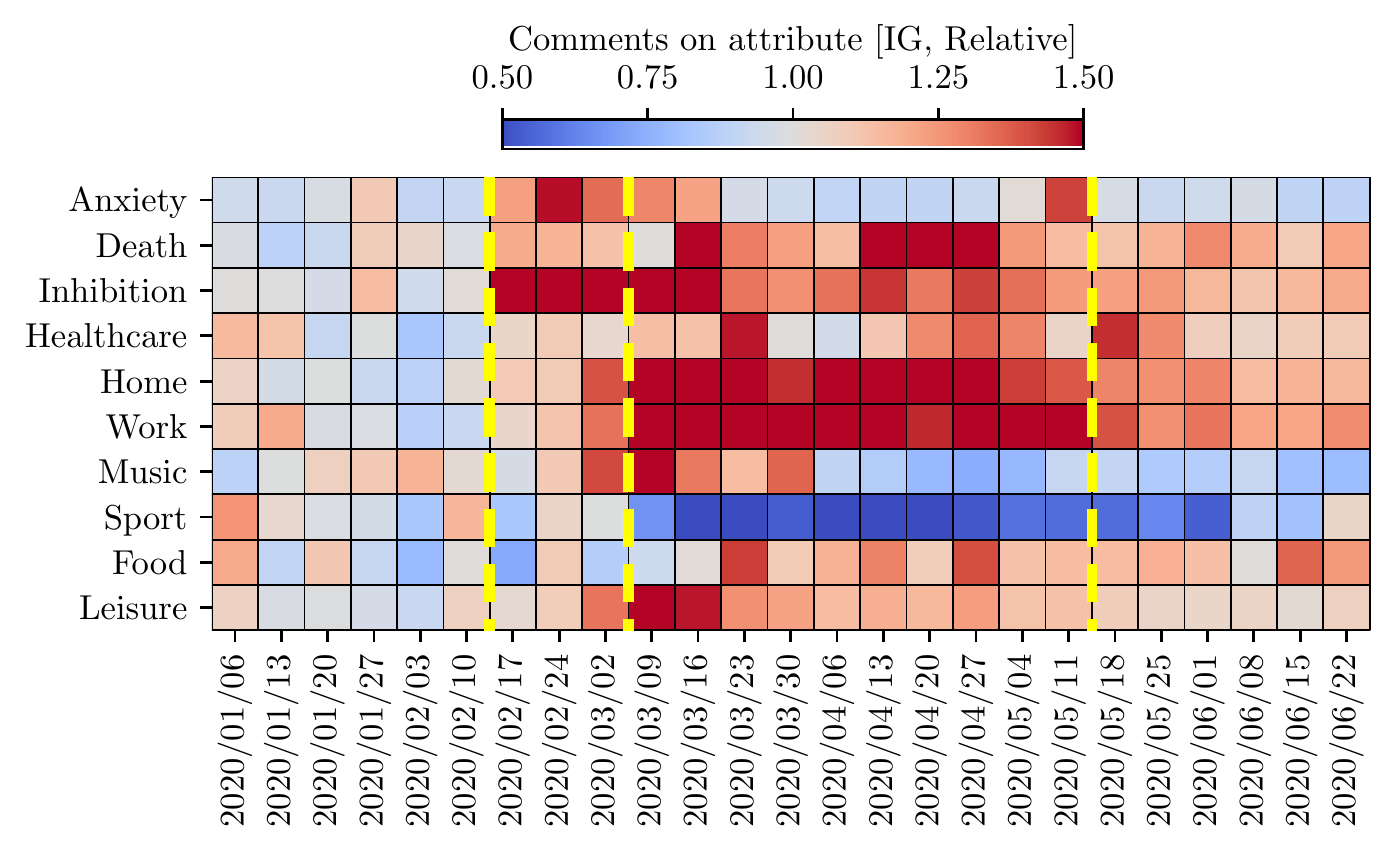}
        \caption{Instagram.}
        \label{fig:liwc_ig}
    \end{subfigure}\\
    \begin{subfigure}{\columnwidth}
        \centering
        \includegraphics[width=0.7\columnwidth]{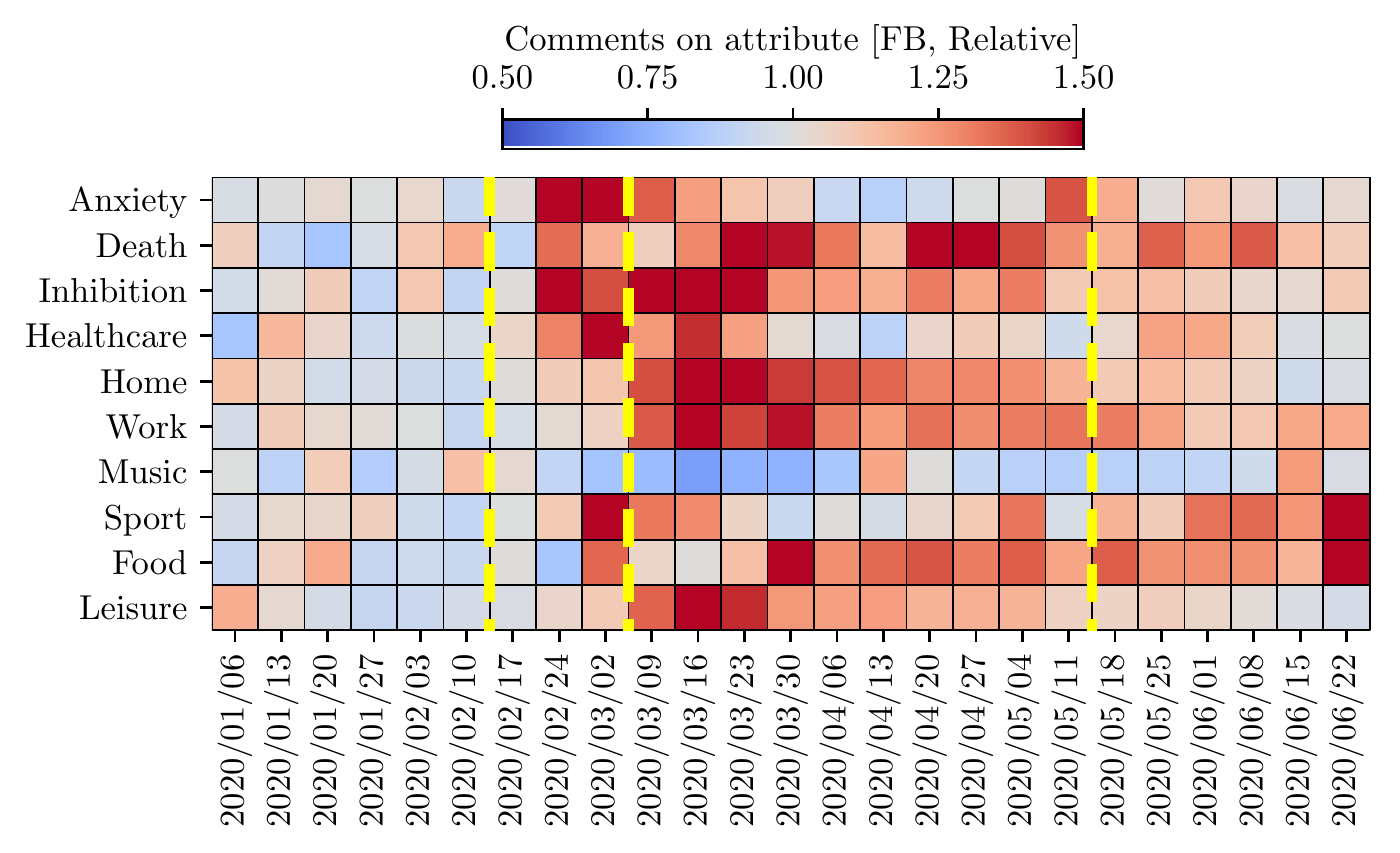}
        \caption{Facebook.}
        \label{fig:liwc_fb}
    \end{subfigure}
    \caption{Relative variation of comments on 10 LIWC attributes, normalized by the pre-lockdown average value.}
    \label{fig:liwc}
\end{figure}

In this section, we focus on the content of comments in terms of their psycholinguistic properties. Our goal is to understand how the users'  topics of conversation shifted during the COVID-19 outbreak. To this end, we use LIWC (see Section~\ref{sec:dataset-methods-topic}). We group LIWC attribute values by week and average the results to measure how the psycholinguistic properties of the debate on the two social networks evolved. We discard those attributes referring to verb tenses and persons and we select those attributes with larger changes, focusing on the analysis on $10$ out of the $83$ attributes. We report the complete set of LIWC attributes in the Appendix. We plot results in Figure~\ref{fig:liwc}, where weeks are arranged on the $x$-axis, and rows represent different LIWC attributes. Vertical yellow lines report the notable events listed in Figure~\ref{fig:events}. Similar to previous analyses, we normalize the values by the pre-lockdown period average value. As such, the average relative variation on the first six weeks is $1$ by definition. Focusing on Instagram (Figure~\ref{fig:liwc_ig}), we notice how attributes related to negative emotions increased during the outbreak. Indeed, \emph{Anxiety}, \emph{Death} and \emph{Inhibition} exhibit an increase between $31\%$ and $52\%$, with some variability across weeks. When the lockdown ends, the comments related to such attributes drop, but they do not return to the pre-lockdown levels (except for \emph{Anxiety}). Similar considerations hold for Facebook (Figure~\ref{fig:liwc_fb}).

Other than these feelings, in Figure~\ref{fig:liwc} we include additional LIWC attributes related to general topics of conversation, limiting to those exhibiting larger variations. We use them to understand how the discussions on the social networks moved. We show general attributes (\emph{Music}, \emph{Sport}, \emph{Food} and \emph{Leisure}) and others related with the pandemic and the lockdown (\emph{Healthcare}, \emph{Home} and \emph{Work}). These last three topics exhibit an increase during the lockdown similar to the feeling attributes. Notice, however, that the increase related to \emph{Healthcare} is smaller than for \emph{Home} and \emph{Work}, hinting that users are prone to discuss their personal activities, rather than debating on the emergency related to COVID itself. Instead, \emph{Music} and \emph{Sport} subdued, the former especially on Facebook and the latter on Instagram. Despite some specific weeks of high activity -- see, for example, the week starting on February 2$^{nd}$ for Sport on Facebook -- this result confirms how the discussions around ordinary conversation topics reduced during the lockdown weeks. We observe an increasing trend of attributes like \emph{Food} and \emph{Leisure}, potentially linked to the shift of interests and activities during the lockdown (see also Section~\ref{sec:4.6}). Finally, some differences hold across profile categories, and, for completeness, we report the per-category breakdown in the Appendix.

\subsection{Discussion around trending topics}
\label{sec:4.6}

\begin{figure}[t]
    \centering
    \includegraphics[width=0.9\linewidth]{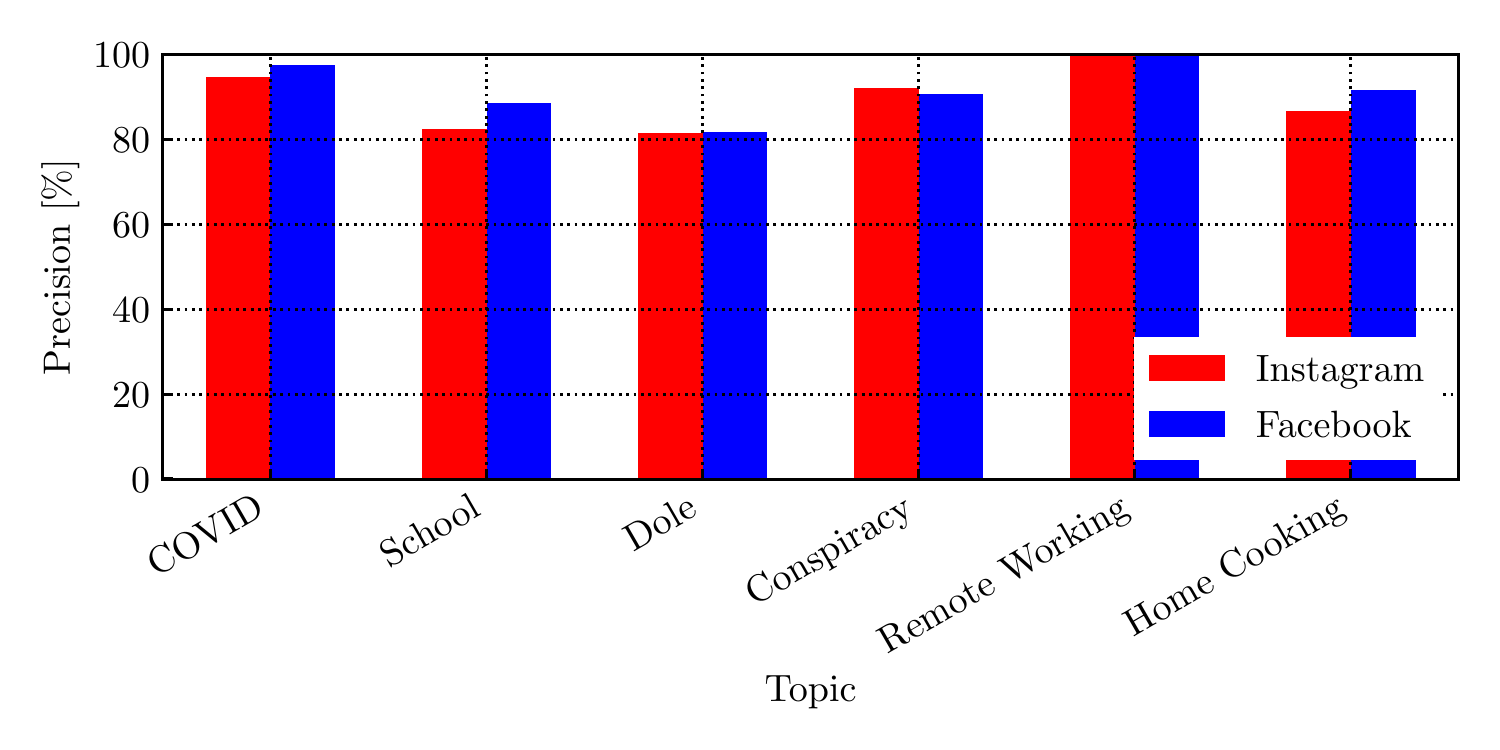}
    \caption{Obtained precision for the chosen topics on the validation set.  \label{fig:true-positives}}
\end{figure}

In this section, we investigate the evolution of some trending topics during the COVID-19 outbreak in Italy. The topics that we target are listed in Section~\ref{sec:dataset-methods-topic}, where we also present our methodology for term-based classification. Since this type of classification technique is prone to errors, we first evaluate the precision for each topic in our dataset, removing the terms that cause frequent misclassification. To this end, we randomly pick a set of $500$ comments per topic for both Instagram and Facebook. We manually label each of these comments as pertinent or not to the assigned topic, i.e., a true or a false positive. We then compute the per-topic per-term precision. From each topic, we remove all the terms generating more false positives than true positives, i.e., having precision lower than 70\%.  For instance, the ``pizza'' term raises more false positives than true positives (40\% precision). This is because, in Italian, this term is frequently used in other contexts, e.g., to indicate frustration. Based on this analysis, we refine our term selection for each topic.\footnote{We make the final list of terms available at~\cite{ourwebsite}.} Then, we randomly pick a new set of $500$ comments per topic on Instagram and Facebook and repeat the manual labeling process to validate the performance. Figure~\ref{fig:true-positives} shows the precision in each topic for Facebook and Instagram, respectively. All topics show consistent performance having similar precision in both social networks. Intuitively, COVID terms are among the most pertinent as very peculiar of this situation. The lowest performance is achieved with Dole, with 81\% precision. This is due to the term ``600 Euro'' (the amount of the Italian Dole), which is sometimes used in other contexts. Finally, we automatically flag about 1 million comments for Facebook and 200 thousand for Instagram with at least one topic. Out of them, only $2.7\%$ and $4.3\%$ of the comments are flagged by multiple topics for Instagram and Facebook, respectively.

\begin{figure}[t]
    \begin{subfigure}{\columnwidth}
        \centering
        \includegraphics[width=0.7\columnwidth]{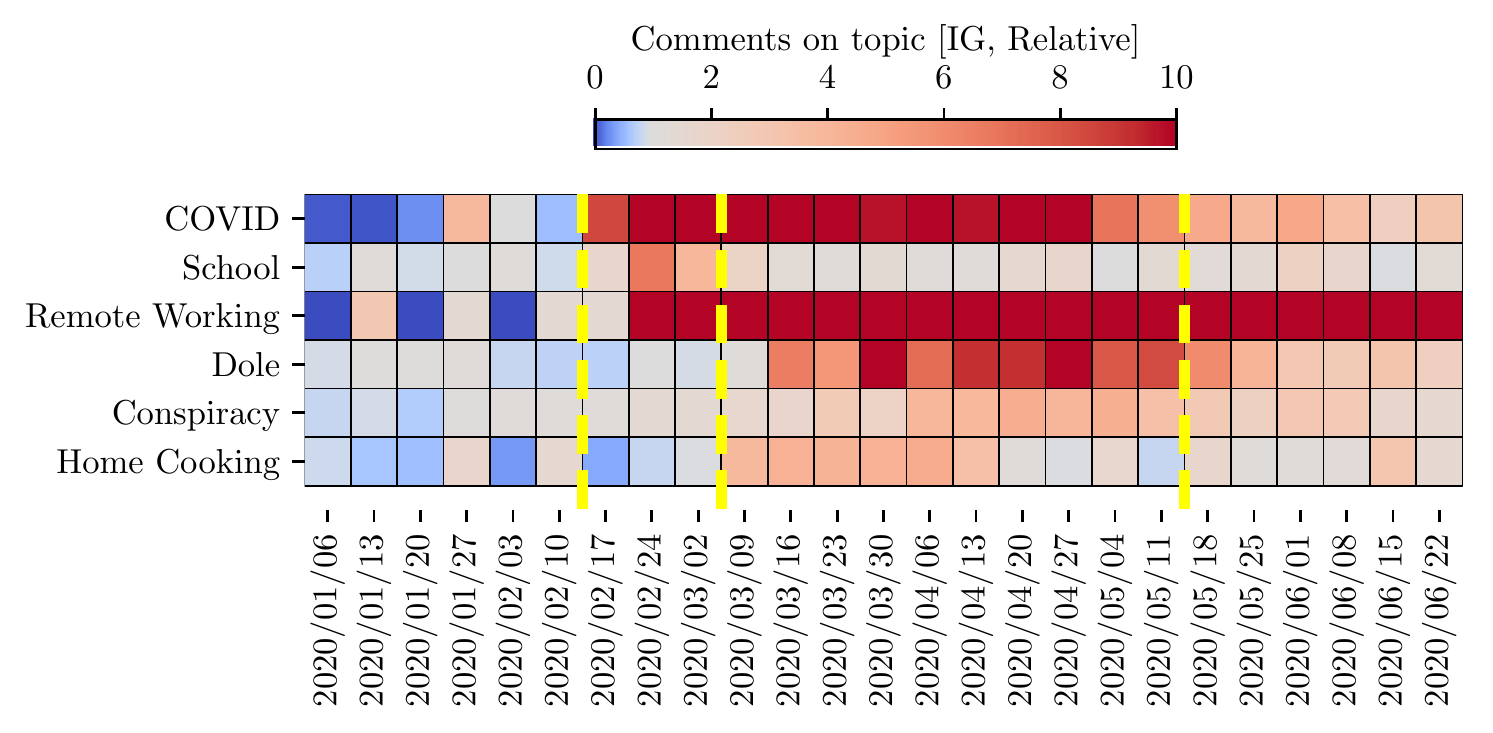}
        \caption{Instagram.}
        \label{fig:topics_ig}
    \end{subfigure}\\
    \begin{subfigure}{\columnwidth}
        \centering
        \includegraphics[width=0.7\columnwidth]{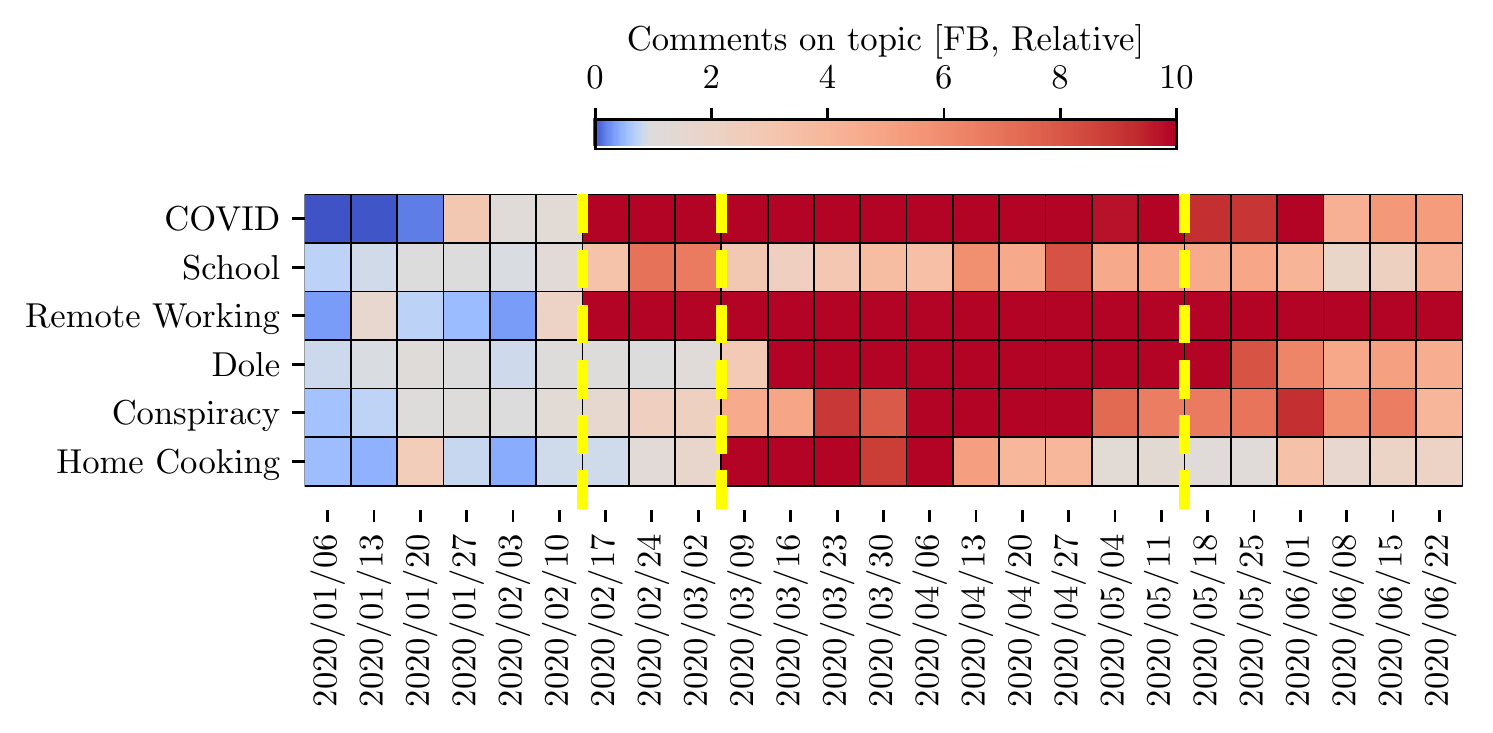}
        \caption{Facebook.}
        \label{fig:topics_fb}
    \end{subfigure}
    \caption{Variation of comments on COVID-19 related topics.}
    \label{fig:topics}
\end{figure}

To study the trending topic evolution, we group the results by week and report the resulting trends in Figure~\ref{fig:topics}. The $x$-axis represents weeks during the whole 6-month period, while topics are arranged on the $y$-axis, and vertical yellow lines report the notable events listed in Figure~\ref{fig:events}. The cell color represents the topic intensity, i.e., the volume of comments on the topic, normalized over its average over the pre-lockdown period. \emph{COVID} topic exhibits the largest increase (more than $10$-times) on both social networks during the lockdown, reaching more than $50$\,k weekly comments. \emph{Remote Working} (more markedly) and \emph{School} (less intensely) gained momentum early as well, suggesting how the discussion on such topics had already started after the first COVID-19 case in Italy (first yellow line). Indeed, the government closed the schools on March 5$^{th}$, and many enterprises had already moved to remote working before the lockdown. Few days after the beginning of the lockdown, the public media started the debate around unemployment support forms. The government issued a \emph{Dole} consisting of a $600$ Euro voucher on March 17$^{th}$, which immediately entered into the discussions on social networks and kept its popularity during all the weeks of the lockdown, with more than $15$\,k comments per week. Interesting is the case of \emph{Conspiracy} that reaches more than $5$\,k weekly comments, especially after the \emph{Bill Gates conspiracy} emerged on the last days of March. Considering \emph{Home Cooking}, we notice how it increased by a factor of $10$ on Facebook and $4$ on Instagram. We later discuss these results in Section~\ref{sec:discusion}.

Focusing on the difference between the two social networks, \emph{Conspiracy} and \emph{Home Cooking} show a higher increase on Facebook than on Instagram. Considering the other $4$ topics, they increase by the same order of magnitude on the two platforms -- with \emph{COVID} and \emph{Remote Working} having the highest absolute increase. When the lockdown ends (third yellow line), most of the topics reduce their popularity. Interestingly, we notice that on Facebook the topics persist longer than on Instagram. See, for example, how \emph{COVID}, \emph{Remote Working} and \emph{Dole} are still popular in late May and early June on Facebook.

To examine more in-depth the users' behavior during the outbreak, we analyze \emph{how} they debate during the lockdown (March 11$^{th}$ to May 18$^{th}$) on each topic using various metrics. For this, we rely on the concept of \emph{commenters} (the users who wrote the comments) and \emph{comments}. Moreover, we consider the comments either as \emph{parent comment}, i.e., direct comment to the influencer post or as \emph{reply} to a previous comment. When not specified, we consider the sum of the two. Based on these notions, we compute the following metrics:
\begin{itemize}
    %About the topic
    \item \emph{Topic share}: the percentage of comments about a topic with respect to the total number of comments. It measures the popularity of the topic.
    \item \emph{Comments per topic}: the absolute number of comments per topic represents the topic's popularity, as the previous metric. It allows us to quantify on how many comments we compute the next metrics. 
    \item \emph{Distinct commenters in \%}: the percentage of commenters posting at least one comment about a topic with respect to the total number of commenters. It measures the spread of the topic among commenters. 
    %About the comment
    \item \emph{Replies in \%}: the percentage of topic comments being replies with respect to the total number of topic comments. This metric measures to what extent the topic is found in debates among commenters. 
    \item \emph{Average replies per comment}: the average number of replies each parent comment on the topic has received. It measures how much a comment on the topic generates debate (even moving to different topics). 
    \item \emph{Active commenters in \%}: the percentage of commenters writing more than one comment about the topic with respect to the number of topic commenters. It measures how much commenters persist on a topic. 
    \item \emph{Average comments per commenter}: the average number of comments each commenter writes on a topic. Similarly to the previous metric, it suggests how much commenters are interested in debating the topic.
    \item \emph{Average comment length}: it gives an intuition of the commenters' interest in arguing their claims with a long text on the topic.
    \item \emph{Comments with link in \%}: the percentage of comments having an external link, for a topic.  Similar to the previous one, it suggests the commenters' interest in corroborating their claims by attaching external resources. 
\end{itemize}

\begin{figure}[!t]
    \begin{subfigure}{\columnwidth}
        \centering
        \includegraphics[width=1\columnwidth]{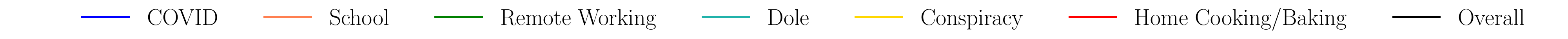}
    \end{subfigure}\\
    \begin{subfigure}{\columnwidth}
        \centering
        \includegraphics[width=1\columnwidth]{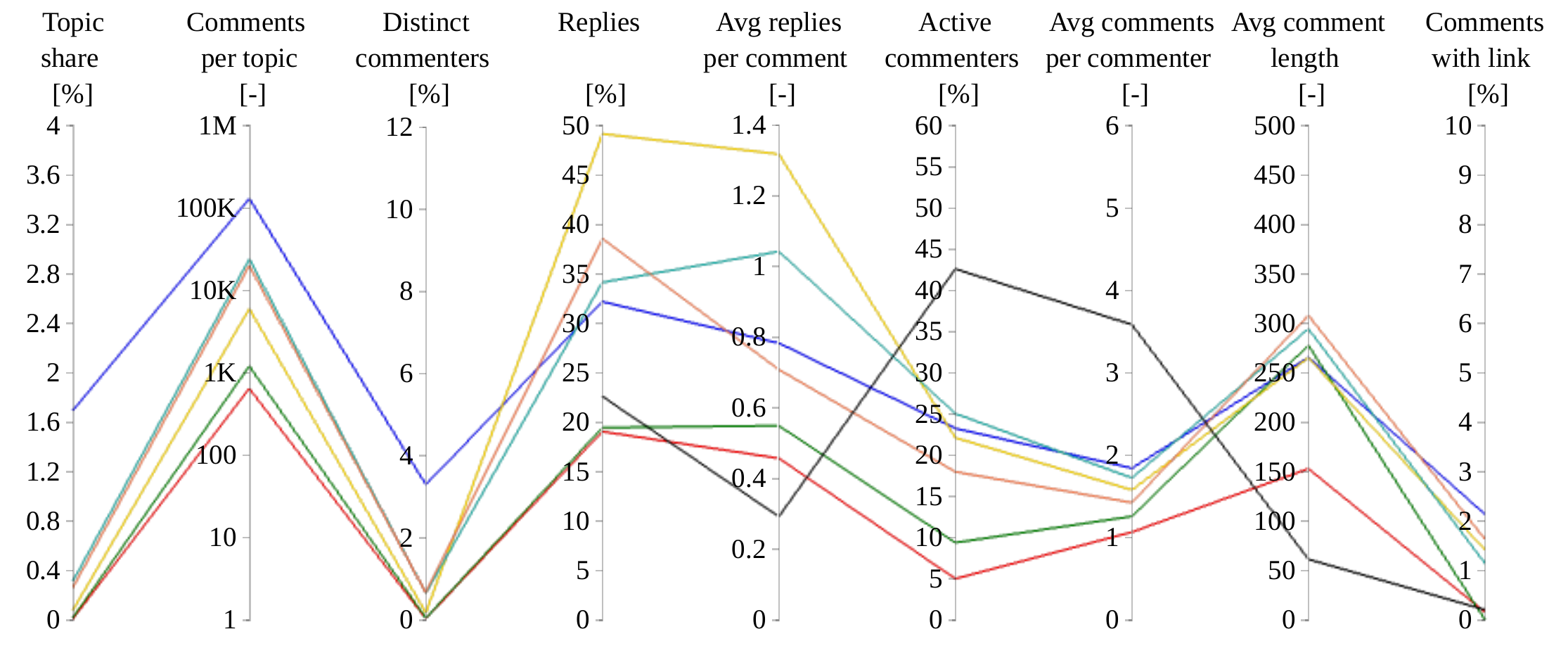}
        \caption{Instagram.}
        \label{fig:parallel_ig}
    \end{subfigure}    
    \begin{subfigure}{\columnwidth}
        \centering
        \includegraphics[width=1\columnwidth]{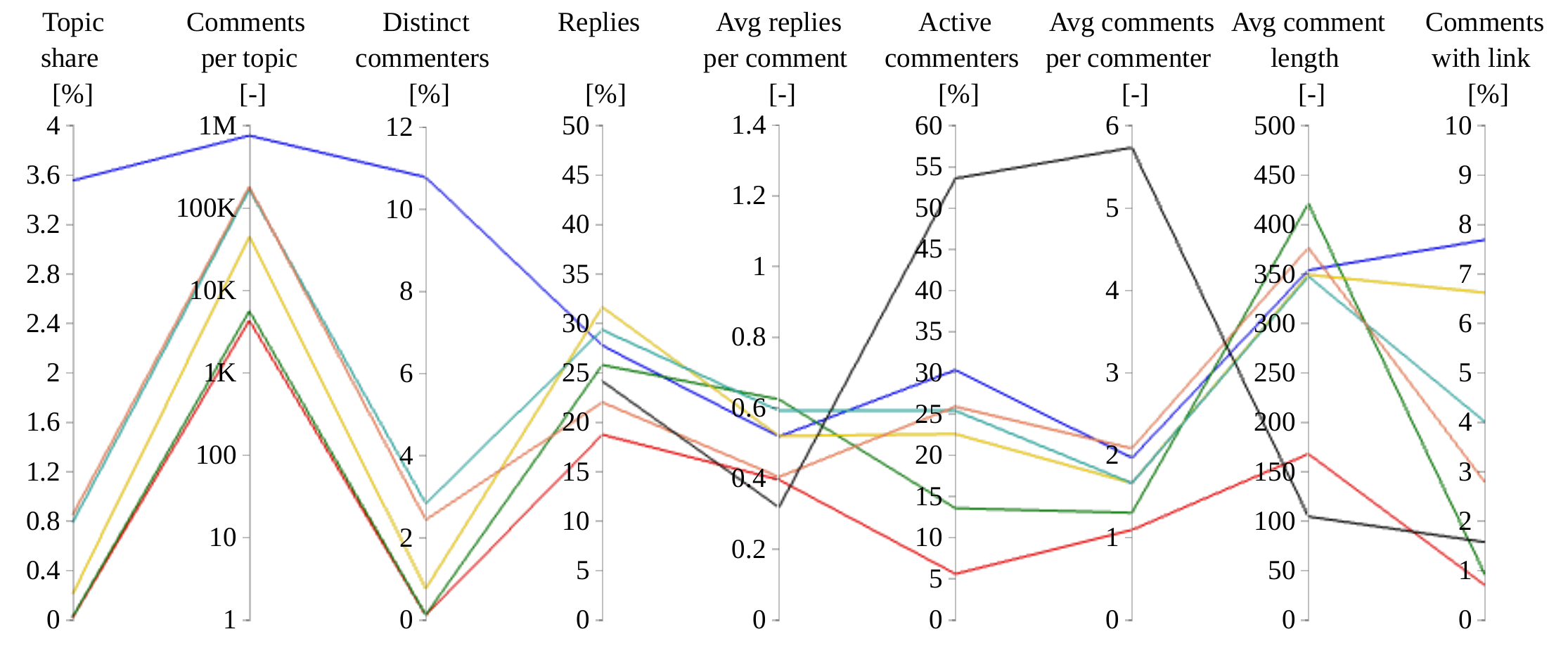}
        \caption{Facebook.}
        \label{fig:parallel_fb}
    \end{subfigure}
    \caption{Statistics of the debate around each topic. The Overall curve reports the behaviour in the whole social network.}
    \label{fig:parallel}
\end{figure}

In Figure~\ref{fig:parallel}, we report a parallel plot for the commenters' behavior of Instagram (top) and Facebook (bottom). The $7$ lines represent the six different topics plus the overall behavior in the \textit{Overall} curve. This curve is computed considering all comments, regardless of the topic. We use it as a reference of what is the usual behavior of the commenters. 

Focusing on the popularity first, we observe how \emph{COVID} is the most popular topic in both social networks. \emph{School} and \emph{Dole} are in second and third position, respectively. Interestingly, the rank on Facebook and Instagram is the same, revealing similar interests in both social networks. In general, trending topics are more popular on Facebook than on Instagram, with the \emph{COVID} topic being more than twice more popular on Facebook than on Instagram. This is also demonstrated by the second axis, in which the \emph{COVID} topic has about 130 thousand comments on Instagram and 750 thousand on Facebook. In both social networks, \emph{Home Cooking} is the least popular one with more than 6 hundred comments on Instagram and 4 thousand comments on Facebook.  Moving to the third column, we study if the popularity of topics is driven by a few commenters generating all the comments (e.g., spammers or flamers), or if many commenters write comments for each topic. This metric follows the previous one for both social networks, with a peak of 11\% of commenters talking about \emph{COVID} on Facebook.  Even the topic with the least distinct commenters on Facebook, i.e., \emph{School}, has the same order of magnitude for distinct commenters (0.24\%) and comment share (0.85\%). These results suggest that a small community of users is responsible for the majority of comments in no case. 

Next, we study the debatability by analyzing how often commenters \emph{reply} to someone else comments. Watching the percentage of topic comments being replies (third column), for Facebook, we do not observe a large deviation from the Overall case (black line). For Instagram, instead, we notice that \emph{Conspiracy} comments are more likely to be replies, i.e., 49\% with respect to only 23\% for Overall behavior. This is also confirmed on the fourth column, representing the average number of replies each comment obtains. Indeed, comments about \emph{Conspiracy} on Instagram are more likely to be within a discussion and also get more replies than other topics. \emph{Dole} gets many replies as well (34\%), hinting that it is another highly debated topic. Instead, on Facebook, all topics receive a higher number of replies than the Overall behavior, but with smaller differences among them.

We now move to evaluate the commenters' persistence in the topics. We measure how likely a commenter writes more than one comment on the same topic (fifth column) and investigate how many comments each commenter writes (sixth column). Again, \emph{Conspiracy} on Instagram is an interesting case, showing high persistence despite the low overall volume (see the first column). Interestingly, the overall behavior highly changes between the two social networks. Indeed, for Facebook, we find a higher percentage of commenters writing more than one comment, and in general, writing $60\%$ more comments than the Instagram counterpart.

Finally, we evaluate the content of comments in the last two columns. In all cases, Facebook comments are longer than Instagram ones. Interestingly, while \emph{COVID} has among the longest comments in both social networks, \emph{Remote Working} shows very different behavior in the two. \emph{Home Cooking} presents the lowest values, suggesting a low level of debate. Regarding the interest in corroborating their comments, Facebook commenters show a higher inclination to include external links. Indeed, more than $8\%$ of comments on \emph{Conspiracy} include a link and $7.6\%$ for \emph{COVID}. Considering the large volume of comments on \emph{COVID}, we find tens of thousands of links.
\section{Discussion and related literature}
\label{sec:discusion}

\subsection{Related work}
\label{sec:relatedwork}

% 1) COVID ON DIFFERENT FIELDS
The COVID-19 pandemic has captured the interest of scientists to understand its impact in different areas.
For example, authors in~\cite{Kraemer493} used mobility data to understand the correlation between mobility and virus transmission in cities across China.
During the lockdown, Internet traffic volume has grown by about $40\%$, sometimes with a decrease of performance, questioning the resiliency of the Internet itself~\cite{cloudflare,fastly}. In our previous work~\cite{FavaleCovid}, we studied the impact of the lockdown on e-Learning systems and networks in our university. 
In the economic field, the impact of the pandemic and the forecast of possible future implications have been studied both from a macro-economic~\cite{Economic1} and from a micro-economic perspective~\cite{EconomicMicro}.
Broadening to social sciences, the changes in food choices following restrictive measures due to COVID-19 has been studied as well~\cite{BRACALE20201423}.

% 2) IMPACT OF OTHER EVENTS ON THE USAGE OF SOCIAL NETWORK
Our study provides investigations on the changes of habits in online social networks during the COVID-19 outbreak.
Other studies in the past followed a similar approach, i.e., collecting and analyzing data from social networks to understand the impact of other events in human history. 
For example, in~\cite{KIM201886} authors studied the interactions of online users on Facebook during disaster responses, with the case study of the 2016 Louisiana flood. Authors in~\cite{PAPAKYRIAKOPOULOS2020100058} studied very active users in the context of political discussions. 
US presidential elections were also widely studied: for example, authors in~\cite{US2016meme} examined how Facebook users used memes to share political ideas during the 2016 elections, and authors in~\cite{Election2012Twitter} studied the tendency in Twitter for individuals to interact with those expressing similar opinions during the 2012 elections. In our previous paper~\cite{Trevisan:2019} and~\cite{Ferreira:2020}, we studied the peculiarity of user interactions with political profiles on Instagram during the Brazilian and European elections (in 2018 and 2019, respectively). Authors in~\cite{KUSEN201837} performed sentiment analysis of the Twitter discussion around the 2016 Austrian elections, showing the difference between the two candidates in terms of sentiment, polarization, likes and retweets. 
Authors of~\cite{WEAVER201918} considered news-sharing on Twitter during the UK general election in 2015, 
showing how news-sharing was affected by the polarization and partisan echo chambers in social media communication. 
Finally, authors in~\cite{OlympicGames} examined social media communities formed during the 2014 Olympic Games and their interactions and kind of exchanged messages. 

% 3) IMPACT OF COVID ON ONLINE SOCIAL NETWORKS
Some papers already analyzed from different perspectives the impact of COVID-19 on online social networks, although it is a recent and ongoing phenomenon. Authors in~\cite{FeldmannIMC} studied network traffic from ISP/IXP vantage points showing how social media application traffic increased in the first weeks of lockdown and then it flattened over time. In~\cite{CorrelationCovidFacebook} authors used aggregated data from Facebook to show that COVID-19 was more likely to spread between regions with stronger social network connections and that this can predict future outbreaks better than physical proximity. Other works explicitly targeted the impact of the pandemic on online social networks. In~\cite{5gConspiracy}, authors analyzed the 5G conspiracy theory in the context of COVID-19 on Twitter, offering suggestions on how to tackle this kind of fake news. The authors of~\cite{perez2020social} studied digital communication on YouTube, Twitter and Instagram during the COVID pandemic in Spain. They found that healthcare professionals and communication media specialized in health largely increased their followers during the lockdown. In this work, we find how Italian politicians largely increased their followers too. In~\cite{bruns2020covid19}, the authors traced the dissemination dynamics of rumors that the pandemic outbreak was somehow related to the rollout of 5G mobile telecommunication technology. They show how the volume of posts on the topic exploded during April 2020, with thousands of links to external websites. In~\cite{celestini2020information}, the authors analyzed the circulation of controversial topics associated with the origin of the virus between Italian users of Facebook. They confirm our findings, showing how content related to the 5G conspiracy emerged during the first phase of the pandemic, together with other controversial topics on migrants and the supposed artificial origin of the virus. The authors of~\cite{zarei2020first} offer a multilingual COVID Instagram dataset that is continuously collected since March 30$^{th}$, 2020. It includes only content obtained monitoring COVID-related keywords, while we provide the complete set of posts and comments for popular Italian influencers. The authors of~\cite{niknam2020covid} identified and analyzed the content of Instagram posts related to COVID-19 in Iran. Similarly to our work, they focus on public accounts and testify the growth of topics related to COVID like stress, fear and government economic support. Finally, the authors of~\cite{cinelli2020covid} analyzed engagement and interest in the COVID-19 topic on Twitter, Instagram, YouTube, Reddit and Gab, finding different volumes of misinformation in each platform. They confirmed how the COVID topic exploded during the pandemic, together with other related topics such as dole and the Bill Gates conspiracy.

\subsection{Discussion on the main findings}

In this paper, we collected a dataset including more than $54$ million comments on over $140$ thousand posts, covering the period surrounding the lockdown and focusing on the top Italian public figures. Despite the size, this dataset has a limited temporal and spatial scope and focuses only on a small fraction of the Italian social media profiles. It does not include any form of private communication or messages between regular users. Hence the findings mainly focus on aggregate behaviour on the posts of the monitored profiles, and we do not track the behaviour of specific users through out the whole platform.

% Quantification
As an alternative to an external social life, we expected the activity on social networks to increase during the lockdown. However, we observed different trends for Facebook and Instagram. On Facebook, we found a large increase during the lockdown weeks in terms of posts ($+47$\%), comments ($+125$\%) and reactions ($+83$\%) -- see Section~\ref{sec:charact} and~\ref{sec:4.1}. Instead, on Instagram, we observed a rather flat trend in terms of posts ($-3$\%) and comments ($+8$\%), while the number of likes almost halved ($-39$\%). We suppose that this is related to the nature of the two social networks. Instagram is based on photos and video and is particularly popular among young people who are keen to share their social life.\footnote{\label{footnote1} Instagram and Facebook demography: \url{https://sproutsocial.com/insights/new-social-media-demographics/}.} Facebook, instead, is frequently used by older people\footnoteref{footnote1} to share comments and opinions and for self-documentation~\cite{alhabash2017tale}.
As a consequence of the restrictions during the lockdown, young people saw a strong decrease of the possibility of social events. 
%dg volendo si potrebbe aggiungere che c'è stata la crescita di piattaforme interessanti per i giovani i.e., tiktok. Ma forse menzionarlo solo qui non ha molto senso. 
Instead, likely older people increased their interactions with topics related to the pandemic, and Facebook appears to be a better platform for this goal.

Particular is the case of politicians. They increased their presence in both social networks and received much more attention (see Section~\ref{sec:4.1}).
%and Figures~\ref{fig:comments_category} and \ref{fig:like_category}). 
Moreover, they acquired a large number of new followers during the lockdown. Indeed, in this period, politicians increased the number of followers almost $4$ times faster compared to the other categories (Figure~\ref{fig:follower}). Similarly to our study, authors of~\cite{haman2020use} showed how political leaders worldwide made a heavy use of Twitter during the pandemic and they significantly increased the number of followers during the pandemic compared to prior months. This was likely driven by the importance and impact of political choices during the lockdown that were affecting day-to-day life. This importance, pushed the population to follow and comment on them more frequently compared to musicians, athletes and entertainers. In turn, these categories had fewer opportunities to interact with followers, as sport events, shows and concerts were halted. The impact of leader responses to COVID and their communication strategy has also been recognized and studied by psychologists~\cite{CovidLeader}. %\mt{Discutere ~\cite{perez2020social}}

% Patterns
We also studied the daily and weekly patterns of usage in Section~\ref{sec:4.4}. As expected, the use of social networks was high during the day, with two peaks after lunch and in the evening, and low during the night. We observed changes due to the pandemic, which sum with the growth of activity previously discussed. There was a $36$\% increase in the morning activity from $7$ AM to $12$ PM and a decrease in the early afternoon and in the evening. Authors of~\cite{cellini2020changes} investigated the use of online media during the Italian lockdown through a survey, and the participants reported an increased use before bedtime, while we witnessed a decrease in the evening regarding online social networks. These variations persisted also after the lockdown when most working activities reopened.  
This suggests a long-term shift in behavior, which could be extremely interesting to monitor in the future. We observed an increase in the activity up to 50\% on Friday and Saturday afternoons and evenings and in the early morning hours of Saturday and Sunday (Figure~\ref{fig:week}). This appears to be linked to the prohibition of in-person social life during the lockdown. Indeed, once the lockdown ended, Friday and Saturday evening activities returned to the pre-lockdown levels, but other changes, such as the higher volume in the morning, persisted.

% Debate
In the first 3-4 weeks of the lockdown, we observed that people were engaged relatively less in discussion with each other, despite the higher number of comments (Section~\ref{sec:4.3}). Subsequently, the amount of discussion increased, reaching and exceeding the pre-lockdown levels on the post-lockdown weeks (Figure~\ref{fig:reply_share}). This is true for both social networks, and it is particularly evident for politicians. We speculate that users were initially less engaged in debating and more prone to support doctors (see \cite{perez2020social}), mourn the victims and look for answers from the influencers and politicians. Instead, after a few weeks, we conjecture that people began discussing about solutions and problems related to the health crisis among themselves, e.g., unemployment. 

In Section~\ref{sec:4.5} we used LIWC to analyze comments, confirming our previous conjecture. We observed that comments expressing negative emotions such as anxiety and inhibition became more popular during the lockdown, especially in the first weeks (see Figure~\ref{fig:liwc}). The authors of~\cite{pellert2020dashboard} found similar trends for Austria, especially regarding anxiety. The authors of~\cite{ASMUNDSON2020102196} showed how fear and anxiety related to COVID-19 were already high in the very initial phase of the outbreak. Moreover, the authors of~\cite{ferrante} performed a survey in Italy during the lockdown, showing that social isolation increased  negative behaviours like alcohol consumption, tobacco smoking and sedentary lifestyle.  In the middle and final weeks, people engaged more on the indirect effects of COVID, such as dole and remote working. Similarly, the authors of~\cite{hung2020social} regarding the US and \cite{niknam2020covid} regarding Iran independently found how the economy, social changes and psychological stress were dominant topics in online social networks during the outbreak. 
We also observed an increasing trend of attributes like Food and Home. We suppose that people, forced to stay at home, shifted their interest to other activities that they could continue with. Unfortunately, the LIWC tool cannot distinguish whether people are talking about themselves or referring to somebody else nor can it detect sarcasm or analogies, which limits further analyses. 
% Topics
In Section~\ref{sec:4.6}, we studied a set of topics related to the pandemic, the lockdown and the social distancing rules in force at that time. We defined these topics by manually constructing bags of words. We analyzed the precision of the chosen words on a validation set. Our study is limited by the manual intervention needed to fill and then check the bag of words. The list of topics is not exhaustive, and while the precision of our methodology is over 80\%, we do not have any control over the recall. Figure~\ref{fig:topics} shows how the COVID topic increased by 10 times in less than 2 weeks, as soon as the lockdown started, and we testify how other topics related to the outbreak, such as Remote Working and Dole, followed similar trends. Other works confirm our findings. The authors of~\cite{yin2020detecting} describe the trending topics in Twitter during the outbreak, finding how \emph{COVID} and related terms such as \emph{death}, \emph{health} and \emph{home} became popular. The authors of~\cite{zhao2020chinese} obtain similar results focusing on Chinese social media. Interestingly, we observed how the discussion around conspiracy theories on the COVID-19 gained momentum, confirming that social networks are a breeding ground for fake news dissemination. This has also been recently shown in~\cite{cinelli2020covid}, with widespread misinformation phenomena occurring during the management of the disease, especially on some social networks.  We notice that, on Facebook, the topics persisted longer than on Instagram, see for example in the debate around \emph{COVID}, \emph{Remote Working} and \emph{Dole}.

Finally, in Section~\ref{sec:4.6}, we also observed that trending topics were debated with diverse user engagement. For example, Figure~\ref{fig:parallel} shows that comments related to all these topics were usually much longer than the average comment, except for Home Cooking. We conjecture that since these topics were surging in popularity and were considered important matters, they needed to be debated more and users needed more extensive explanations of their claims. Comments about conspiracies were the ones that obtained the most replies. Moreover, they were also the ones that most frequently came with a hyperlink. We hypothesize that these theories were discredited by many people, and hence long discussions and arguments appeared in these social networks, with also many users referencing external news websites, either to confirm or reject the related conspiracy. Indeed, the authors of~\cite{moscadelli2020fake} testify how links to fake news articles circulated heavily on social media during the COVID-19 outbreak.

\section{Conclusion}
\label{sec:conclusions}

The COVID-19 pandemic impacted the healthcare systems of the hit countries and many aspects of our society and habits. Studying how people reacted to such a historical event is, thus, of paramount importance for understanding human behavior, and social media provide the researchers with a unique lens for this goal. 

Using a large dataset collected on Instagram and Facebook around popular Italian public figures, we observed an increase of social network usage during the lockdown. Especially on Facebook, we testified a general growth in the volume of posts ($+47$\%), comments ($+125$\%) and reactions ($+83$\%). The debate level decreased on the first phases of the lockdown but then exceeded the pre-lockdown levels on the post-lockdown weeks. As a result of the lockdown, Italian users were more active in the morning and on Friday and Saturday evenings. We also found that political profiles, differently from the others, acquired a large number of new followers on Instagram. During the lockdown, comments expressing concern, such as anxiety and inhibition, increased. The discussion around COVID-19 soared during February 2020, and we testified how other topics related to the outbreak, such as Remote Working and Dole, followed similar trends. On Facebook, these topics generally persisted longer than on Instagram. The debate around ordinary topics of conversation (sport and music) diminished.

We believe that our results provide many insights on the use of social networks and its evolution at the time of COVID-19. We hope our work fosters further research in the psychological and sociological areas. To this end, we release an anonymized version of our dataset \cite{ourwebsite} to allow the reproducibility of our results and the use of these data in other contexts.

\section*{Acknowledgements}
The research leading to these results has been funded by the Smart\-Data\-@Po\-li\-TO center for Big Data technologies. We also thank the staff of \url{https://www.pubblicodelirio.it/} for helping us in building the profile lists and the English experts of Politecnico di Torino for their precious help on the revision of the text.

\small
\bibliographystyle{ieeetr}
\bibliography{reference}

\newpage

\section*{Appendix}
\noindent
We report additional results that have not been included in the paper for the sake of brevity. Some of the material here presented may be useful for the readers that are willing to dig more in depth into our findings.

\subsection*{Posts per profile}
\begin{figure}[h!]
    \centering
    \begin{subfigure}{.45\linewidth}
        \includegraphics[width=1.\linewidth]{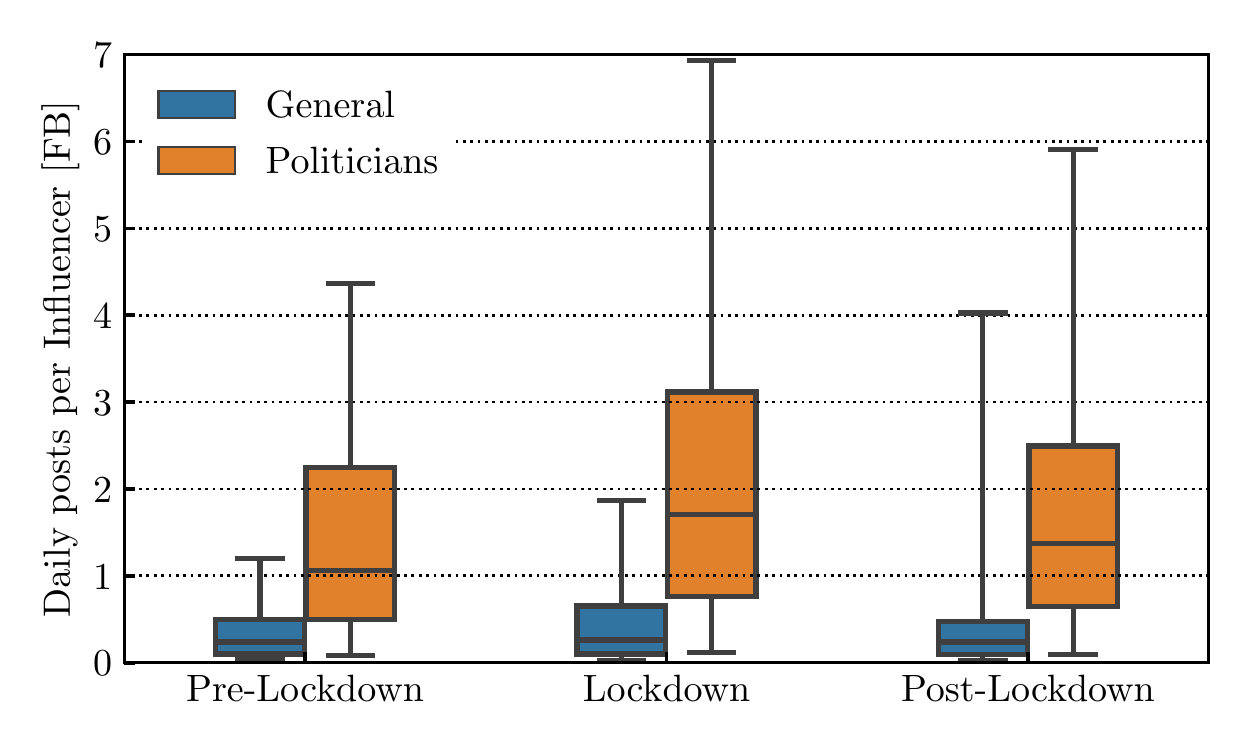}
        \caption{Facebook}
        \label{fig:posts_per_capita_fb}
    \end{subfigure}
    \begin{subfigure}{.45\linewidth}
        \includegraphics[width=1.\linewidth]{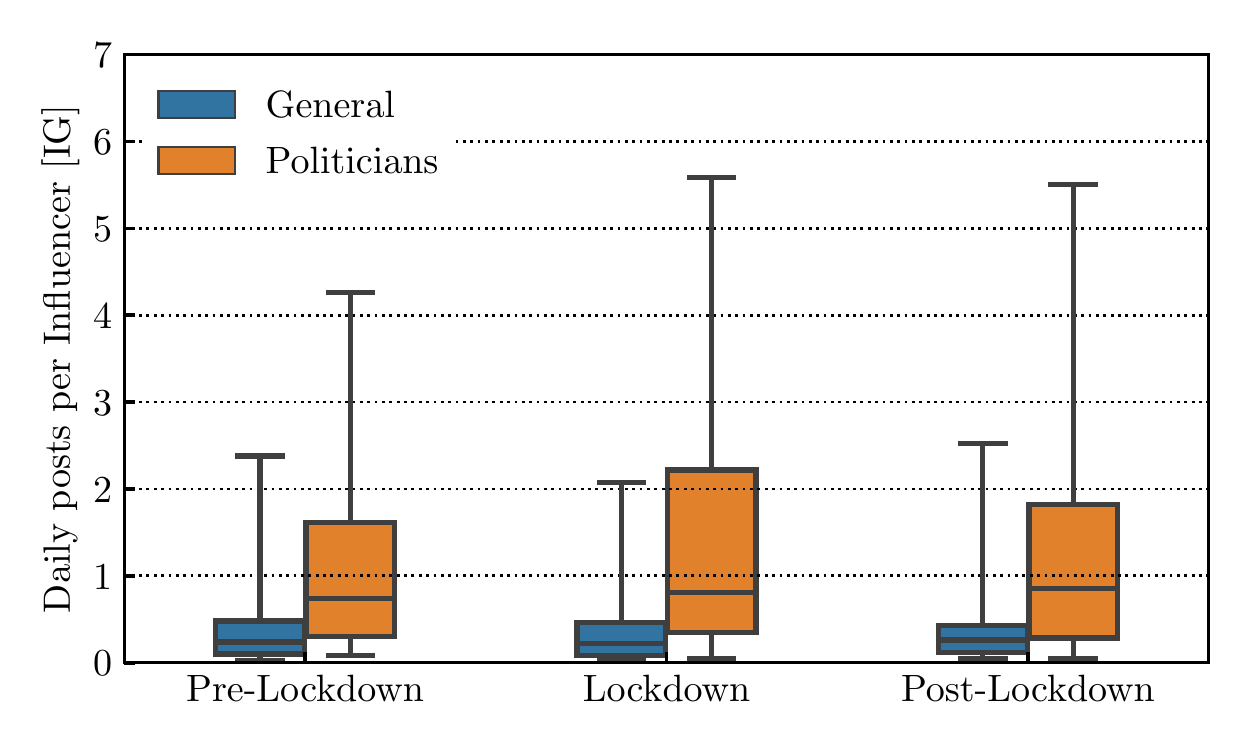}
        \caption{Instagram}
        \label{fig:posts_per_capita_ig}
    \end{subfigure}
    \caption{Daily number of posts per influencer.}
    \label{fig:posts_per_capita}
\end{figure}

\noindent
In Figure~\ref{fig:posts_per_capita}, we show the distribution of daily posts per profile, comparing the three periods defined in Table~\ref{fig:events}. We find that Politicians  publish more posts, on both social networks, with a peak during the lockdown. No notable trend emerges for the other categories (here reported aggregated).

\subsection*{Reactions to posts with different length}
\begin{figure}[h!]
    \centering
    \includegraphics[width=.6\linewidth]{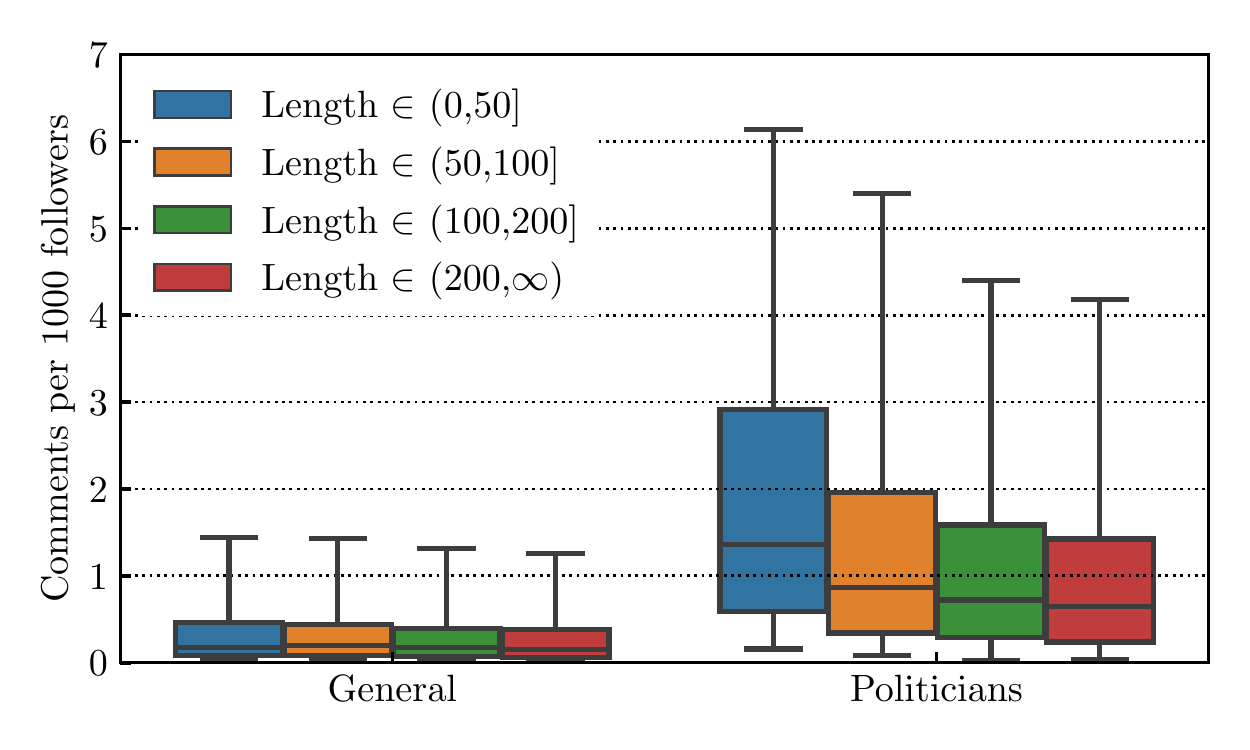}
    \caption{Comment reactions to posts with different length.}
    \label{fig:comm_post_len}
\end{figure}

\noindent
We show in Figure~\ref{fig:comm_post_len} the distribution of the number of comments that posts with different length received. We grouped posts into 4 groups depending on the length of their text/caption. The bounds of the groups are chosen so as to obtain roughly the same number of posts in each. We then show the distribution of the number of comments they receive. For a fair comparison, we normalize the number of comments by the number of followers each profile has at the time the post was created.

For Politicians, longer posts receive fewer comments on average (see median values represented by a black stroke). For General profiles, we do not observe such a phenomenon, and the amount of comments they receive seems not to be affected by the post length. The figure also confirms the findings of our previous work~\cite{Trevisan:2019}, showing that Politicians on average obtain relatively more comments than other categories.

%\newpage
\subsection*{Follower increase over a wider time span}
\begin{figure}[h!]
    \centering
    \centering
    \includegraphics[width=.55\linewidth]{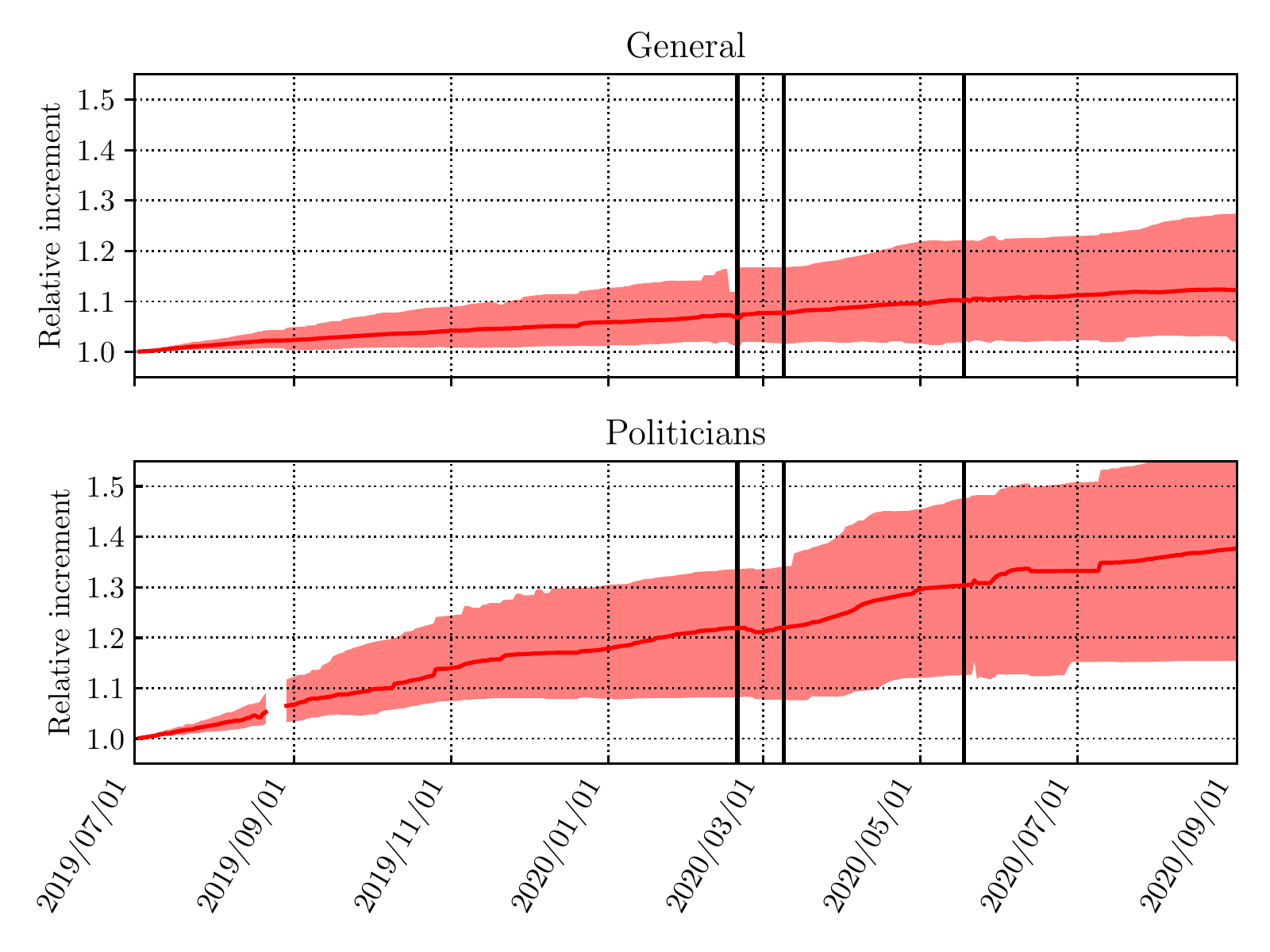}
    \caption{Follower variation over 14 months. 25$^{th}$, 50$^{th}$ and 75$^{th}$ percentiles are shown.}
    \label{fig:follower_long}
\end{figure}

\noindent
In Figure~\ref{fig:follower_long}, we extend the analysis of Figure~\ref{fig:follower} to 14 months around the lockdown period (from July 2019 to August 2020). The figure confirms our findings, and show that Politicians acquired new followers at a higher rate during the lockdown.

\subsection*{Weekly pattern during the post-lockdown}

\begin{figure}[h!]
    \centering
    \centering
    \includegraphics[width=.65\linewidth]{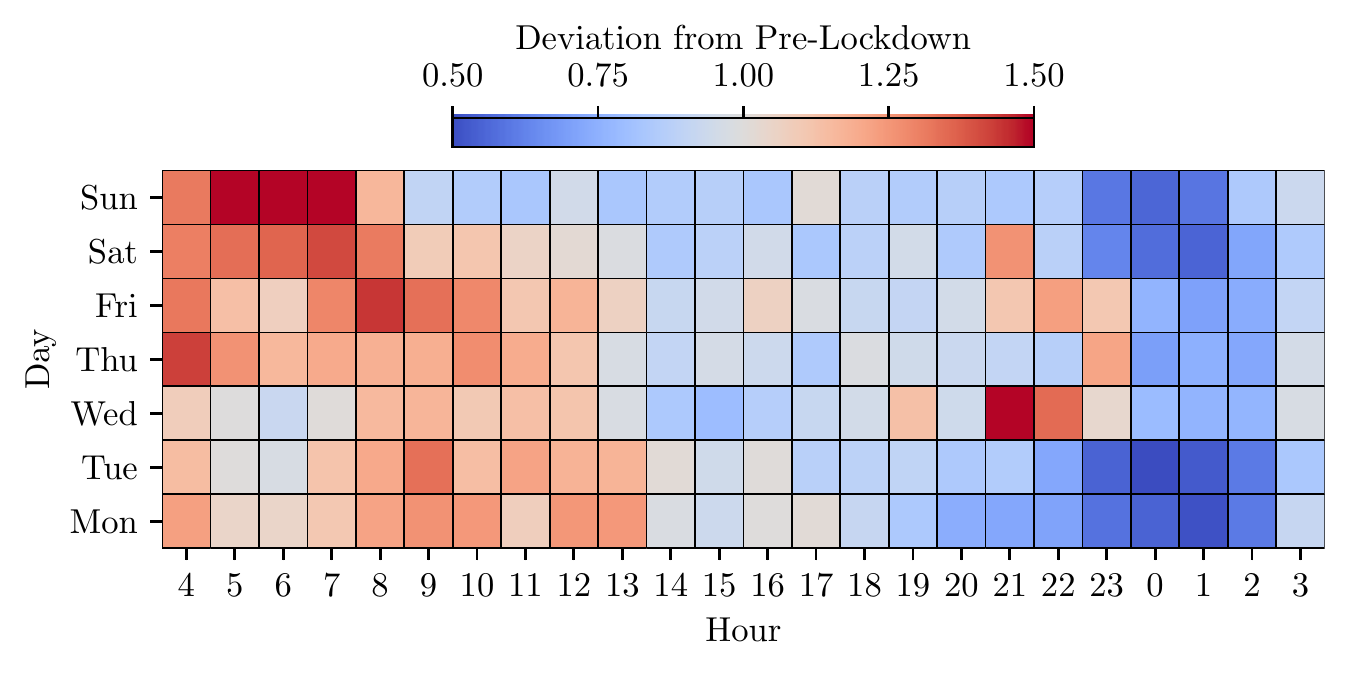}
    \caption{Variation of the weekly pattern of Instagram comments in the post-lockdown compared to the pre-lockdown period. Notice that the x-axis begins at 4 AM.}
    \label{fig:deviation_after}
\end{figure}

\noindent
In Figure~\ref{fig:deviation_after}, we show the deviation (increase or decrease) in number of comments of the post-lockdown period compared to the pre-lockdown, computed as the ratio between the two shares (compare it with Figure~\ref{fig:week}). Some of the pattern variations observed during the lockdown attenuate, or disappear, while others hold also during the post-lockdown, e.g., the morning increase.

\newpage

\subsection*{LIWC attributes by profile category}
\begin{figure}[h!]
    \centering
    \begin{subfigure}{.4\linewidth}
        \includegraphics[width=1.\linewidth]{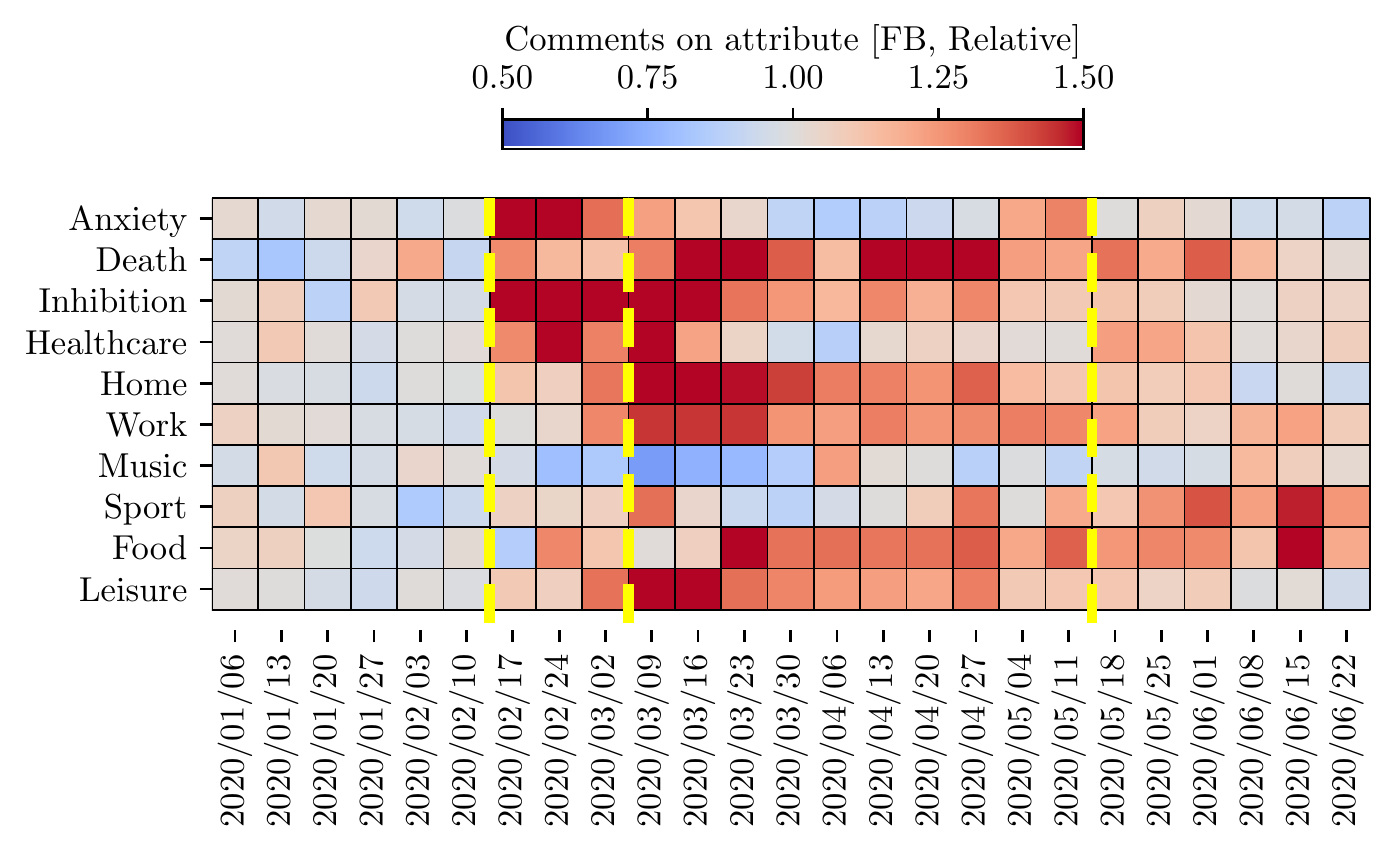}
        \caption{Politicians}
    \end{subfigure}
    \begin{subfigure}{.4\linewidth}
        \includegraphics[width=1.\linewidth]{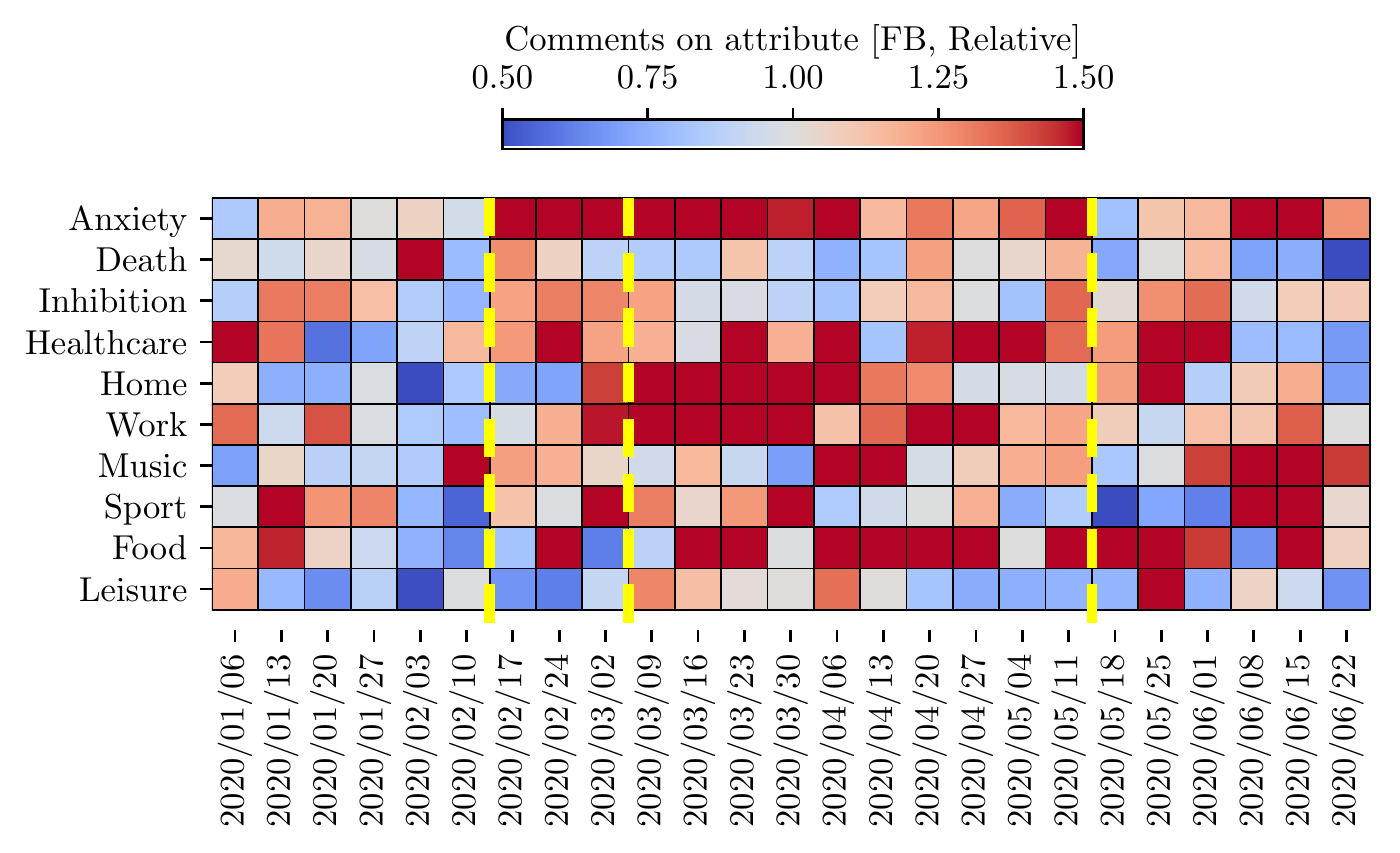}
        \caption{Musicians}
    \end{subfigure}
    \begin{subfigure}{.4\linewidth}
        \includegraphics[width=1.\linewidth]{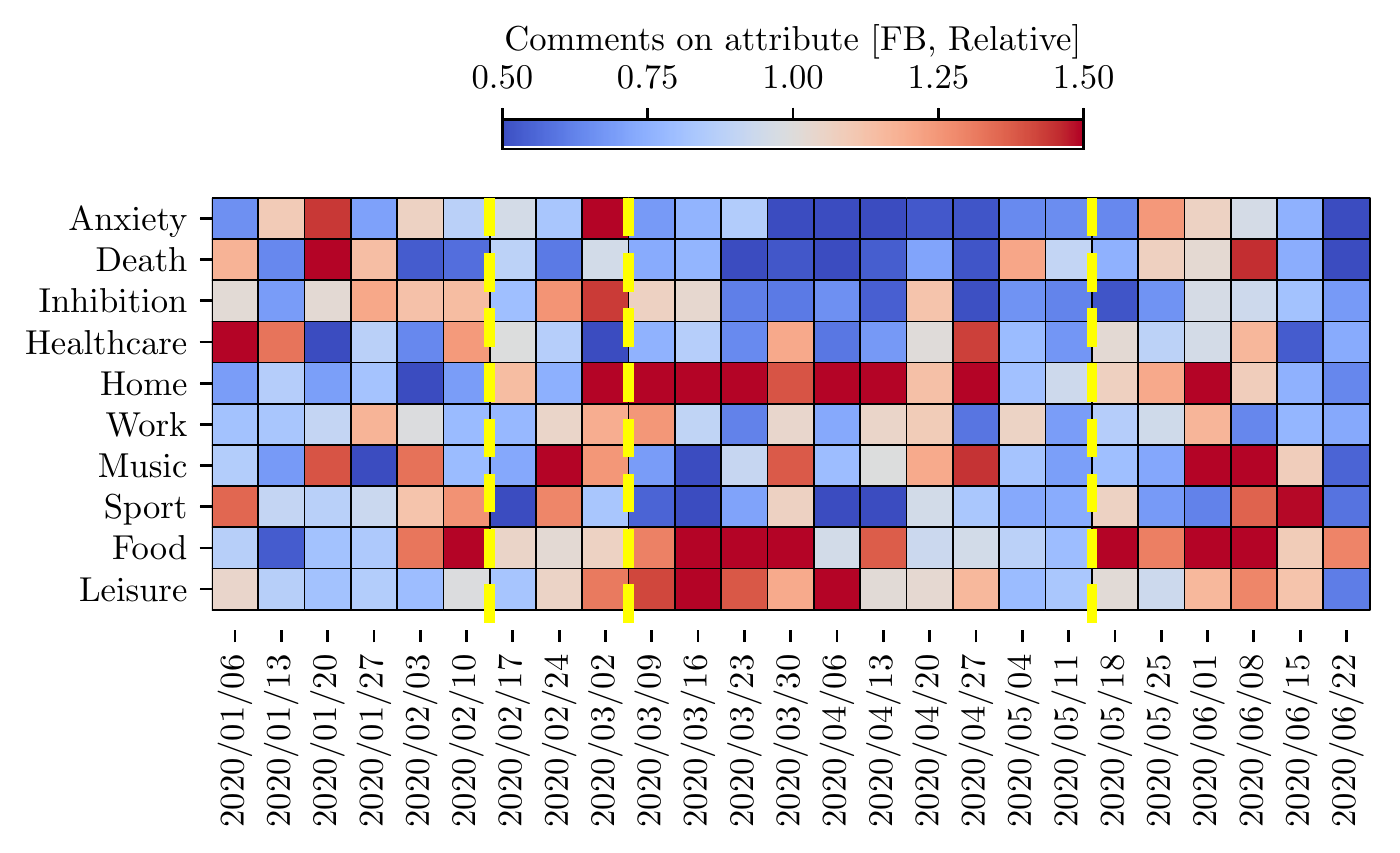}
        \caption{Athletes}
    \end{subfigure}
    \begin{subfigure}{.4\linewidth}
        \includegraphics[width=1.\linewidth]{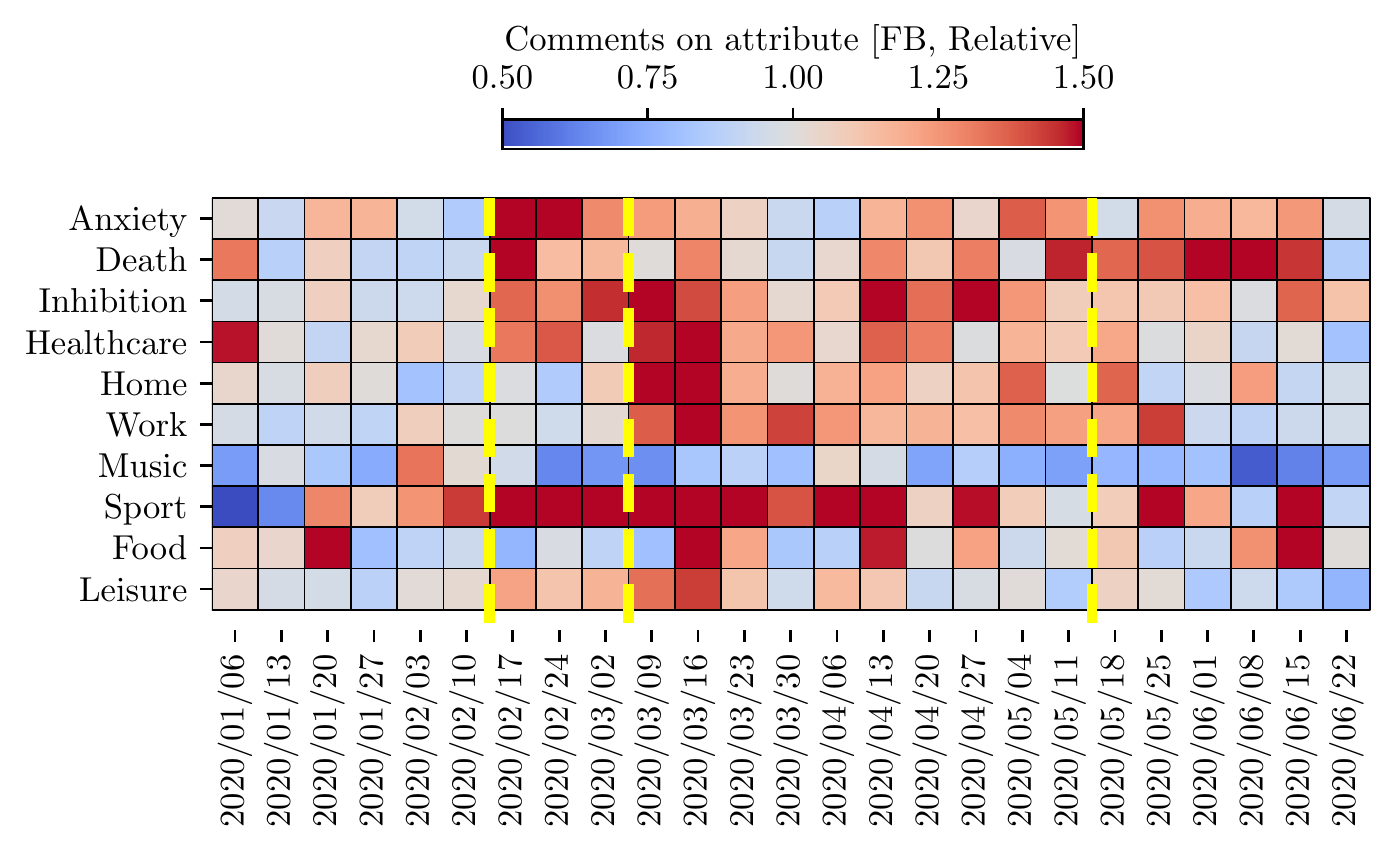}
        \caption{Entertainers}
    \end{subfigure}
    \caption{Trend of selected LIWC by category (Facebook).}
    \label{fig:liwc_fb_cat}
\end{figure}

\begin{figure} [h!]
    \centering
    \begin{subfigure}{.4\linewidth}
        \includegraphics[width=1.\linewidth]{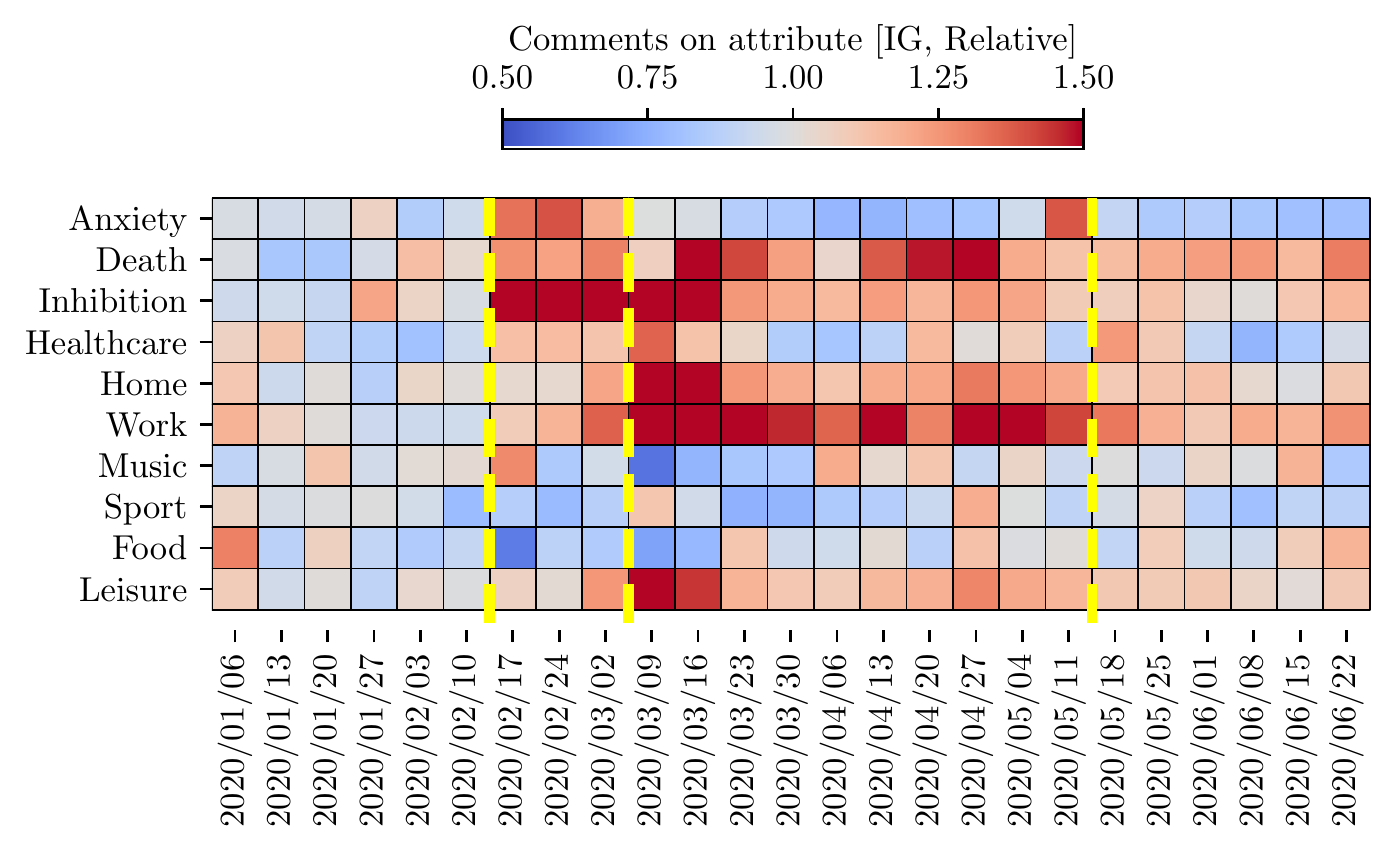}
        \caption{Politicians}
    \end{subfigure}
    \begin{subfigure}{.4\linewidth}
        \includegraphics[width=1.\linewidth]{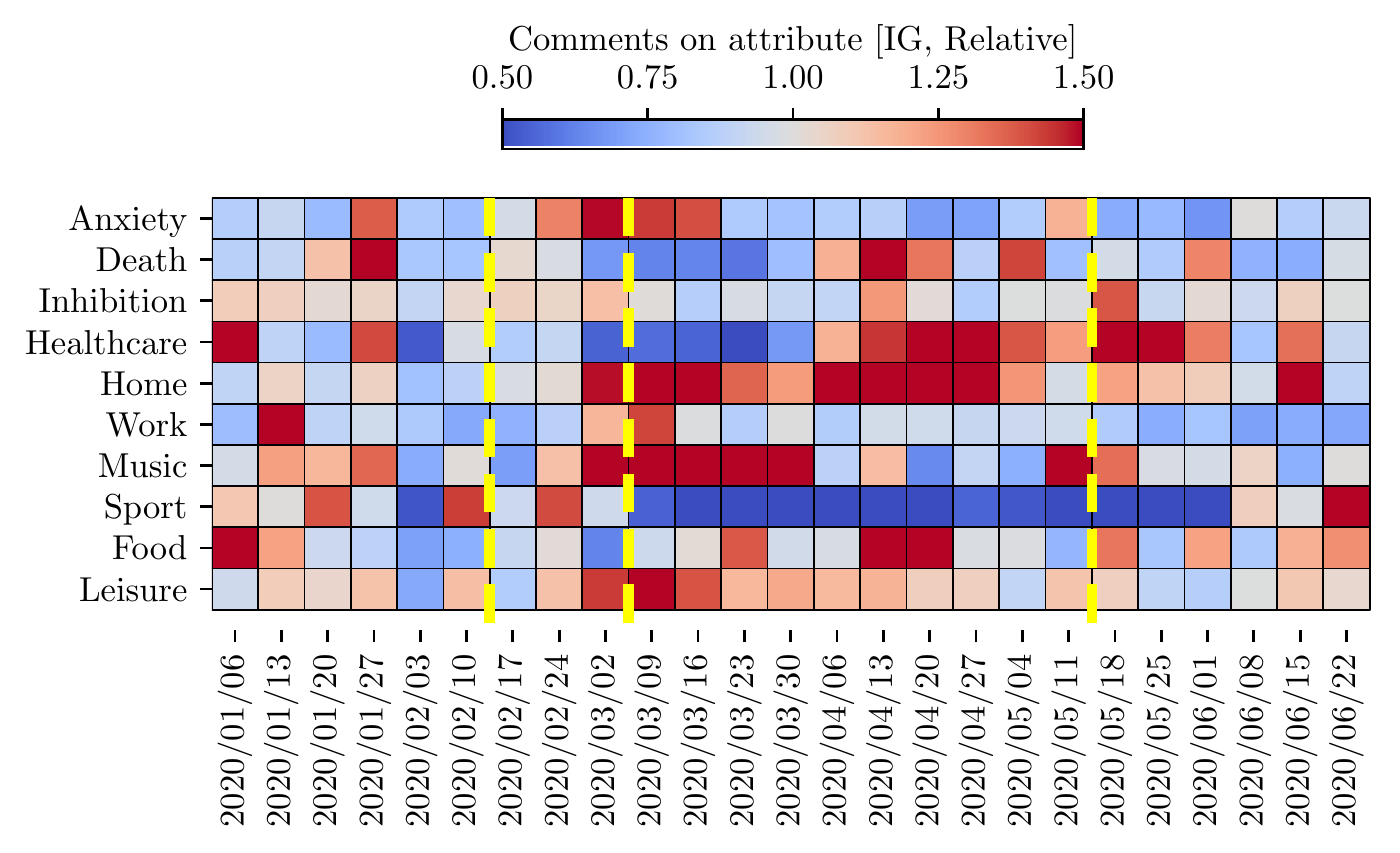}
        \caption{Musicians}
    \end{subfigure}
    \begin{subfigure}{.4\linewidth}
        \includegraphics[width=1.\linewidth]{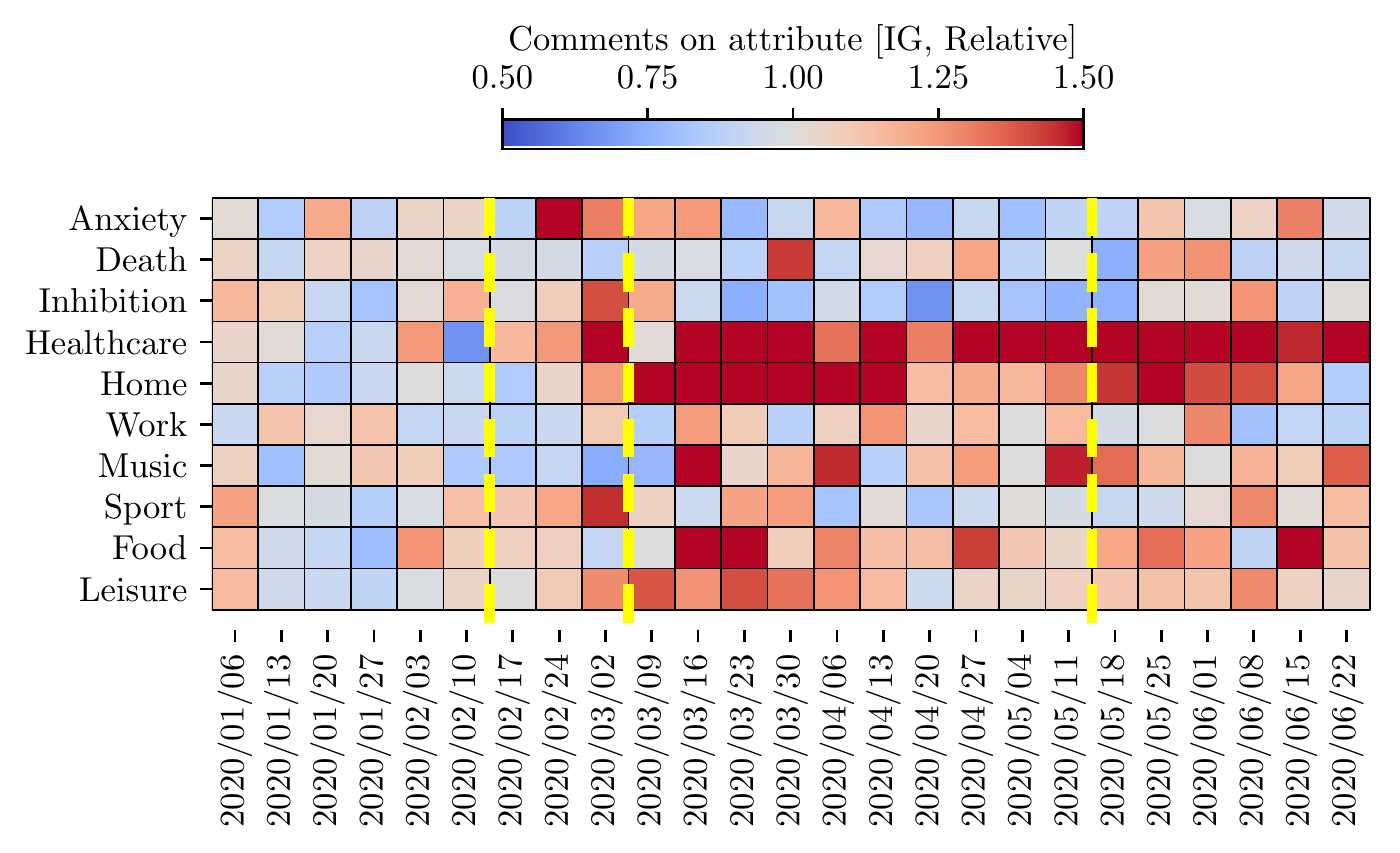}
        \caption{Athletes}
    \end{subfigure}
    \begin{subfigure}{.4\linewidth}
        \includegraphics[width=1.\linewidth]{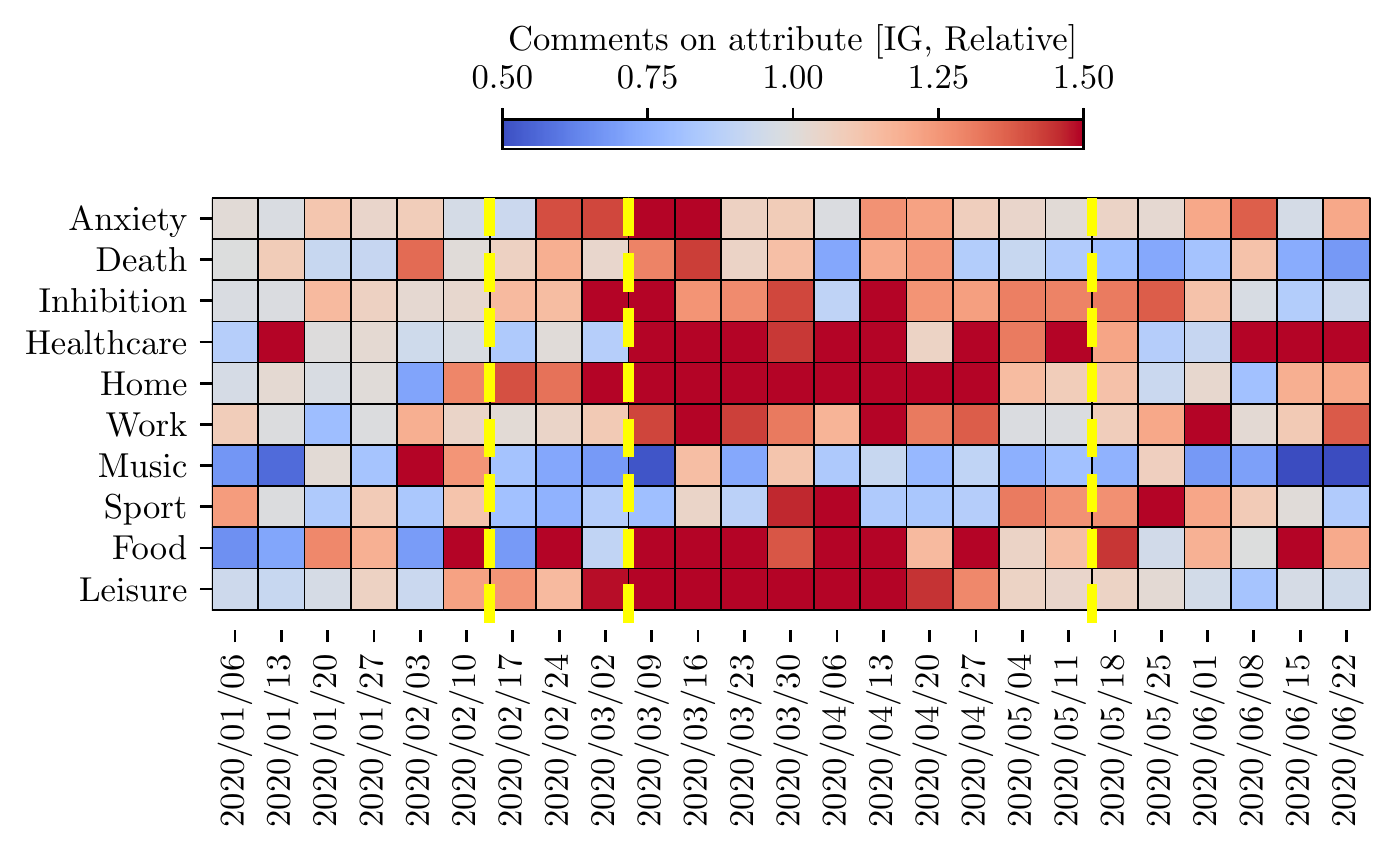}
        \caption{Entertainers}
    \end{subfigure}
    \caption{Trend of selected LIWC by category (Instagram).}
    \label{fig:liwc_ig_cat}
\end{figure}

\noindent
We here breakdown the analysis presented in Section~\ref{sec:4.4}, providing in Figure~\ref{fig:liwc_fb_cat} and Figure~\ref{fig:liwc_ig_cat} the trends of the selected LIWC attributes separately by profile category. The presented pictures show a high degree of diversity, and dissimilarities emerge. While for some attributes (e.g., Home or Food) the increase appears generalized, in other cases (concern attributes, Work and Leisure) the picture is variegated.

\clearpage
\subsection*{Trends for all LIWC attributes}
\begin{figure}[h!]
    \centering
    \begin{subfigure}{.49\linewidth}
        \includegraphics[width=1.\linewidth]{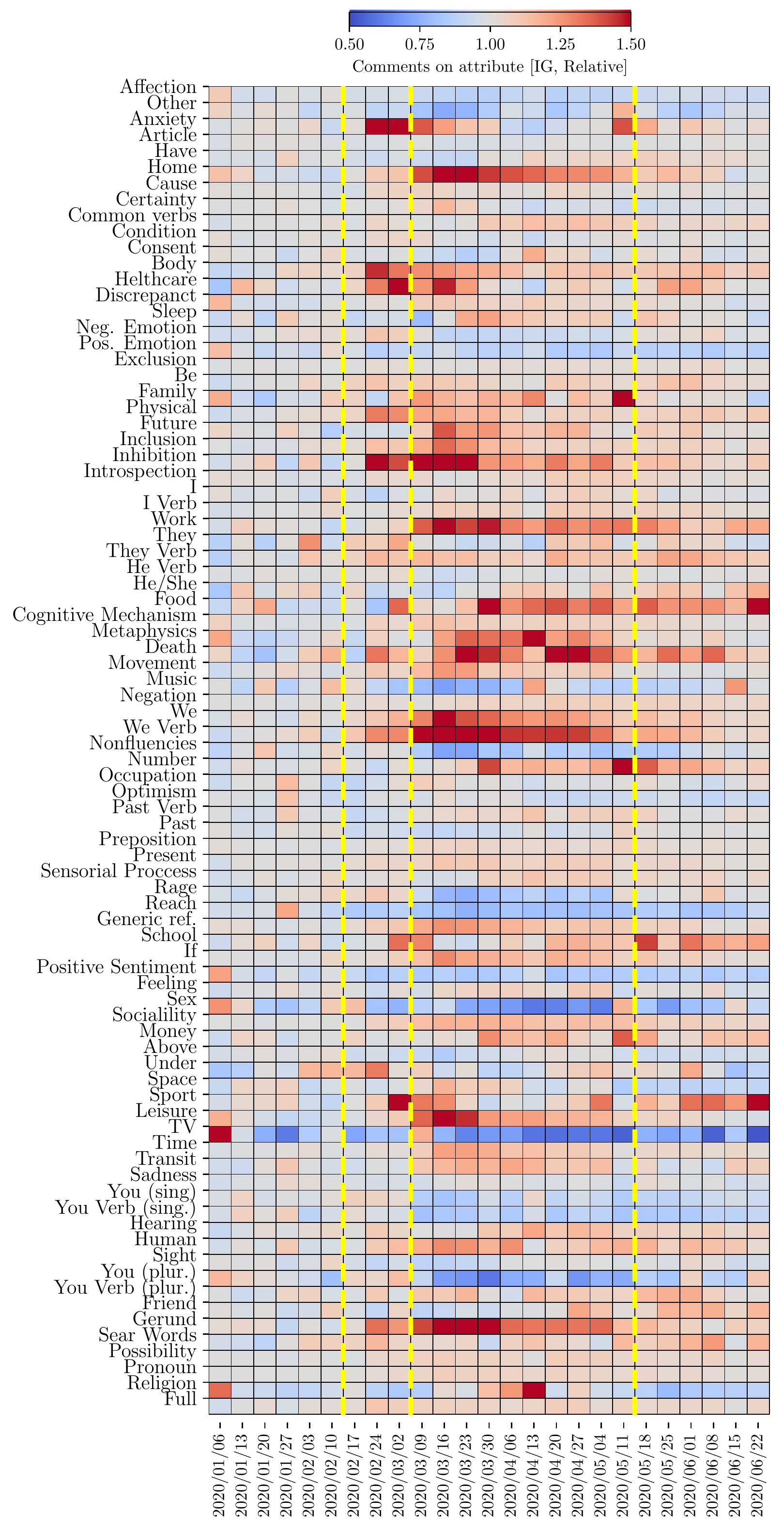}
        \caption{Facebook}
        \label{fig:posts_per_capita_fb}
    \end{subfigure}
    \begin{subfigure}{.49\linewidth}
        \includegraphics[width=1.\linewidth]{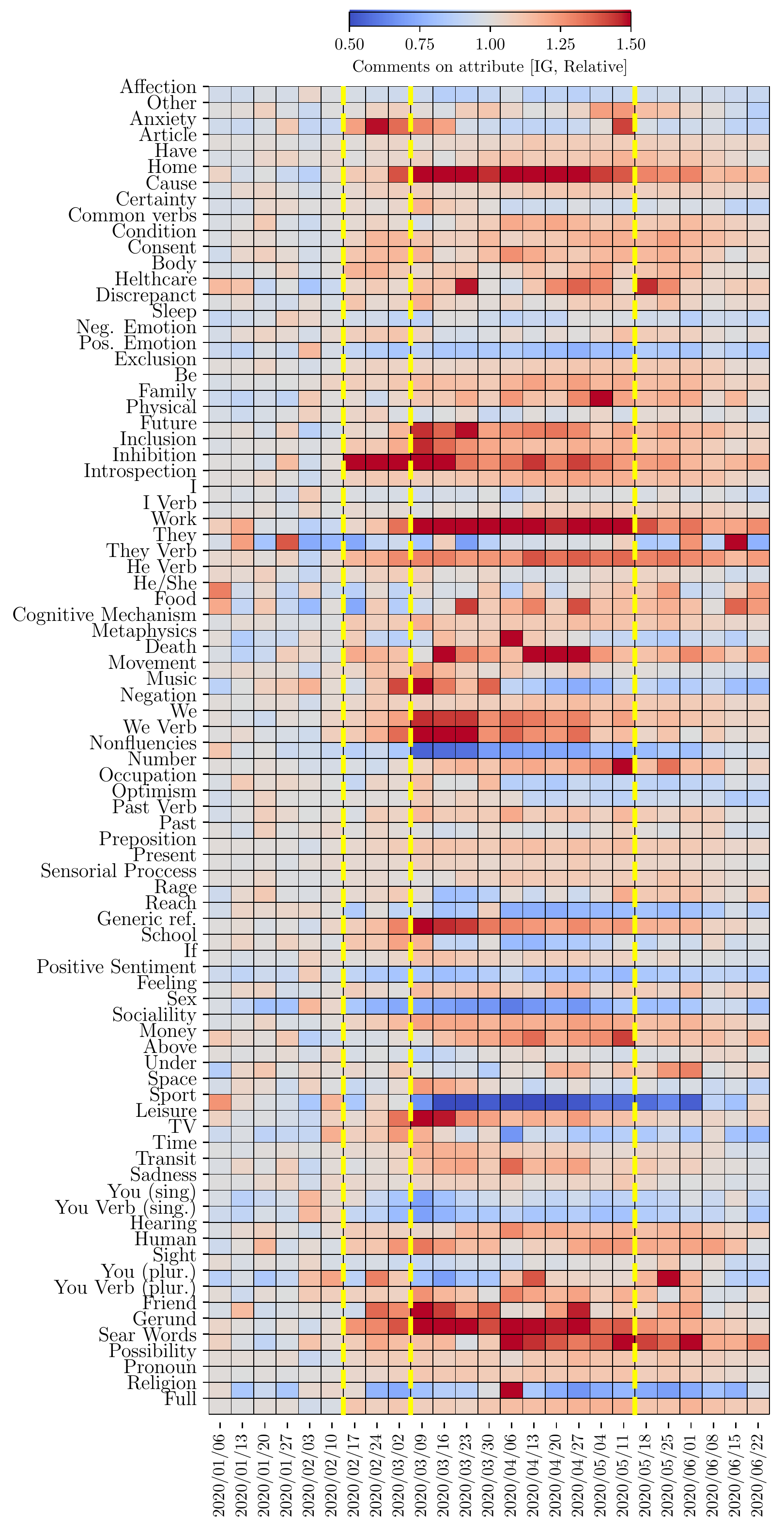}
        \caption{Instagram}
        \label{fig:posts_per_capita_ig}
    \end{subfigure}
    \caption{Trend of \emph{all} LIWC attributes over time.}
    \label{fig:liwc_all}
\end{figure}

\noindent
In our analysis on the psycholinguistic properties of comments (Section~\ref{sec:4.4}), we selected only a subset of the available attributes in LIWC. %We have chosen those attributes exhibiting the most pronounced trends, after excluding those related to verb tenses or verb persons. 
In Figure~\ref{fig:liwc_all}, we provide the complete picture, showing the trends for all the $83$ LIWC attributes, in order to provide a complete view.

\end{document}